\documentclass[5p]{elsarticle}

\usepackage{graphicx}
\usepackage{amssymb}
\usepackage{amsmath}
\usepackage{gensymb}
\usepackage{textcomp}
\usepackage{color}
\allowdisplaybreaks
\usepackage[hyphens]{url}
\usepackage{nomencl}
\usepackage{framed}
\usepackage{etoolbox}
\usepackage{setspace}

\usepackage{lineno}
\usepackage{float}
\usepackage[utf8]{inputenc}
\usepackage[T1]{fontenc}

\makenomenclature

\begin{document}

\begin{frontmatter}


\title{The worm-LBM, an algorithm for a high number of propagation directions on a lattice Boltzmann grid: the case of phonon transport}

\author{Ren\'e Hammer\cormark[cor1], Verena Fritz, Natalia Bedoya-Mart\'{i}nez\cormark[cor1]}
\cortext[cor1]{Corresponding authors: Rene.Hammer@mcl.at, \\ OlgaNatalia.Bedoya-Martinez@mcl.at}
\address{Materials Center Leoben Forschung GmbH, Roseggerstrasse 12, 8700 Leoben, Austria}

\begin{abstract}
The lattice Boltzmann method (LBM) is a numerical approach to tackle problems described by a Boltzmann type-equation, where time, space, and velocities are discretized to describe scattering and advection. Even though the LBM executes advection along a lattice direction without numerical error, its usage in the high Knudsen number regime (ballistic) has been hindered by the ray effect problem (for dimensions greater than 1D). This problem has its origin in the low number of available propagation directions on standard LBM lattices. Here, to overcome this limitation, we propose the worm-lattice Boltzmann method (worm-LBM), which allows a high number of lattice directions by alternating in time the basic directions described within the next neighbor schemes. 
Additionally, to overcome the velocity anisotropy issue, which otherwise clearly manifests itself in the ballistic regime (e.g. the $\sqrt 2$  higher grid velocity of the  D2Q8 scheme along the diagonal direction compared to the axial one), the time-adaptive scheme (TAS) is proposed. The TAS method makes use of pausing advection on the grid, allowing to impose not only isotropic propagation but also arbitrary direction-dependent grid velocity. Last but not least,  we propose  a grid-mean free path (grid-MFP) correction to correctly handle the aforementioned velocity issue in the diffusive limit, without affecting the ballistic one.  We provide a detailed description of the TAS method and the worm-LBM algorithm, and verify their numerical accuracy by using several transient diffusive-ballistic phonon transport cases, including different initial and boundary conditions.
 We demonstrate the accuracy of the new worm-LBM to describe problems where a high angular resolution (i.e. a high number of propagation directions) is required, as the in-plane thermal transport problem under adiabatic-diffusive boundary conditions. In this particular case, we show that schemes with a low number of propagation directions (D2Q8) result in an overestimation of the analytical Fuchs-Sondheimer solution for intermediate and high Knudsen numbers, and that schemes with a higher number of propagation directions are required to correctly describe the problem.   Overall, the new, very accurate, and efficient worm-LBM algorithm, free of  numerical smearing and false scattering, has the potential to be at the forefront of the numerical solvers to tackle the advective part of different equations in a wide field of applications.  
\end{abstract}

\end{frontmatter}

\renewcommand\nomgroup[1]{%
  \item[\bfseries
  \ifstrequal{#1}{G}{Greek Symobls}{%
  \ifstrequal{#1}{L}{Latin Symbols}{%
  \ifstrequal{#1}{V}{Vectors}{}}}%
]}
\setlength{\nomitemsep}{-\parsep}

\nomenclature[L]{$f$}{phonon distribution function}
\nomenclature[L]{$f_\text{BE}^{eq}$}{Bose-Einstein distribution}
\nomenclature[V]{$\vec{x}$}{position}
\nomenclature[L]{$L$}{domain length}
\nomenclature[L]{$x_\text{cum}$}{cumulative distance in direction i}
\nomenclature[L]{$x_\text{ax}$}{cumulative distance along axial direction}
\nomenclature[L]{$x_0$}{x-position initial Gaussian}
\nomenclature[L]{$y_0$}{y-position initial Gaussian}
\nomenclature[L]{$X$}{normalized x-position}
\nomenclature[L]{$Y$}{normalized y-position}
\nomenclature[L]{$N_x$}{number of grid points in x-direction}
\nomenclature[L]{$N_y$}{number of grid points in y-direction}
\nomenclature[G]{$\xi$}{normalized time}
\nomenclature[G]{$\Theta$}{normalized temperature}
\nomenclature[L]{$Q_x$}{normalized heat flux in x-direction}
\nomenclature[L]{$Q_y$}{normalized heat flux in y-direction}
\nomenclature[G]{$\sigma_0$}{width of initial Gaussian}
\nomenclature[V]{$\Delta\vec{x}$}{grid vector}
\nomenclature[L]{$t$}{time}
\nomenclature[G]{$\Delta t$}{time step}
\nomenclature[L]{$T$}{temperature}
\nomenclature[L]{$T_\text{c}$}{temperature cold boundary}
\nomenclature[L]{$T_\text{h}$}{temperature hot boundary}
\nomenclature[V]{$\vec{q}$}{heat flux vector}
\nomenclature[V]{$\vec{k}$}{phonon wave vector}
\nomenclature[L]{$k$}{phonon wave number}
\nomenclature[L]{$p$}{phonon polarization}
\nomenclature[G]{$\omega$}{phonon frequency}
\nomenclature[G]{$\omega_0$}{gray model phonon frequency}
\nomenclature[V]{$\vec{v}_\text{g}$}{phonon group velocity}
\nomenclature[G]{$\lambda$}{phonon mean free path}
\nomenclature[G]{$\kappa$}{thermal conductivity}
\nomenclature[G]{$\kappa_{\text{effective}}$}{apparent effective thermal conductivity}
\nomenclature[G]{$\alpha$}{thermal diffusivity}
\nomenclature[V]{$\hat{s}$}{direction}
\nomenclature[G]{$\theta$}{polar angle}
\nomenclature[G]{$\theta_i$}{discretized angle}
\nomenclature[G]{$\phi$}{azimutal angle}
\nomenclature[G]{$\beta$}{angle between direction and projection axis}
\nomenclature[G]{$\tau$}{phonon relaxation time}
\nomenclature[L]{$e$}{energy density}
\nomenclature[L]{$e_i$}{discretized directional energy density}
\nomenclature[L]{$e^\text{eq}$}{total equilibrium energy density}
\nomenclature[L]{$e^\text{eq}_i$}{directional equilibrium energy density}
\nomenclature[V]{$\vec{c_i}$}{grid velocity vector}
\nomenclature[L]{$D$}{phonon density of states}
\nomenclature[G]{$\Omega$}{solid angle}
\nomenclature[L]{$N$}{number of atoms}
\nomenclature[L]{$V$}{unit-cell volume}
\nomenclature[L]{$Q$}{number of directions}
\nomenclature[L]{$w_i$}{angular weighting factor}
\nomenclature[L]{$i$}{index discretized directions}
\nomenclature[L]{$j$}{discrete space index x-direction}
\nomenclature[L]{$k$}{discrete space index y-direction}
\nomenclature[L]{$n$}{discrete time index}
\nomenclature[L]{$\text{Kn}$}{Knudsen number}
\nomenclature[L]{$W$}{weighting factor}
\nomenclature[L]{$W_\text{corr}$}{weighting factor corrected}
\nomenclature[G]{$\Lambda$}{grid mean free path}
\nomenclature[L]{$q_\text{corr}$}{velocity projection correction}
\nomenclature[G]{$\Lambda_\text{corr}$}{grid mean free path correction}
\nomenclature[G]{$\delta$}{distance from desired ray}
\nomenclature[L]{$R$}{radius circular initial condition}
\nomenclature[L]{$r$}{distance from midpoint}

\section{Introduction}
\label{S:1}

\textit{“There is no such thing as a perfect advection scheme - only differing degrees of badness”} \cite{Gresho1998}. The fundamental reason for this is that in any numerical scheme that discretizes space and time, advection does not propagate the field from one lattice site to another but rather interpolates it. This generally leads to numerical dissipation and/or dispersion errors, also known as numerical smearing and angular false scattering.  There are many strategies to lower the  dispersion error of advection dominated problems within real space schemes  \cite{Ewing2001,Leonard1991,Liu2015,Hou2012}. However, a perfect reproduction of the analytical dispersion of the underlying differential equation is only achieved for the "magic ratio" between time step and grid spacing having a Courant-Friedrichs-Lewy (CFL) number equal to one. This  can in general only be fulfilled in 1D (see  1D \cite{Hammer2014a} vs. 3D \cite{Hammer2014b}). A rather inexpensive discretization method that overcomes these problems  is the lattice Boltzmann method (LBM). The Lagrangian discretization of the LBM couples velocity and real space exactly. Hence, the CFL number can be chosen to be one for all lattice directions, and advection is reproduced exactly  along the lattice.  From a practical point of view, an advantage of the LBM is that the discretized field variable (e.g. density) is updated locally in time and space. This makes the LBM efficient and highly parallelizable, such that massive parallel computing architectures with minimal communication requirements and overhead can be used \cite{OBRECHT2013252,CALORE20161}. 
The LBM has been widely used in fluid dynamics, radiative transfer, neutron transport, rarefied gas dynamics, and phase change \cite{succi2001,McHardy-16,Cen-2020,Bindra2012,Zhang2005,Li2018,He2019}. Moreover, it is used in wave propagation, electrodynamics, quantum mechanics, relativistic fluids, ion and electron transport,  and diffusive-ballistic phonon transport \cite{Succi2015,Hauser2019,Zhong2006,Jiang2017,Chattopadhyay2014,Escobar2008,GUO20161}. The last one, relevant for thermal management at the nano- and micro-scale, 
is in the focus of the present work.

\begin{framed}
\printnomenclature
\end{framed}

At length scales comparable to the mean free path of the thermal energy carriers (phonons in the case of semiconductors), or at time scales shorter than their relaxation times, the diffusive picture of heat transport has to be replaced by the so-called ballistic transport ~\cite{Siemens-2010,Minnich2011}. Phonon transport in the ballistic regime is relevant in  fields as varied as nano- and micro-electronics, thermal barriers, energy harvesting, nano- and opto-mechanics and quantum technologies \cite{Cahill2014,Volz2016}.
In this regime, as long as the particle nature of the heat carriers prevails, the Boltzmann transport equation (BTE) must be used. The BTE can be used to describe both the continuum (diffusive) and sub-continuum (ballistic) regime. It is routinely used in combination with density functional theory (DFT) calculations to describe bulk phonon transport \cite{Broido2007}. However, to  account for interface and surface scattering leading to geometrical size effects in thermal transport, it is necessary to  solve the time- and space-dependent BTE. This is computationally demanding as it involves seven independent dimensional variables  accounting for space, time, and velocity. Monte Carlo (MC) methods have been used traditionally to tackle the high dimensionality issue. However, they are less effective for thermal phonon transport studies, wherein most of the cases one deals with problems close to equilibrium, and small temperature differences have to be tackled accurately. In this case, MC methods spent too much time sampling the equilibrium distribution. This issue has been overcome with the energy-based variance-reduced Monte Carlo formulation \cite{Peraud-2011}. This method is highly efficient, if a linearization of the BTE is justified, and if the superposition principle can be used to propagate the computational particles independently of each other through the system. Nonetheless, it suffers from statistical noise as any MC based technique. For treating temperature-dependent scattering, large thermal gradients, non-equilibrium situations, and in general for obtaining deterministic results, a direct discretization (spatial and angular) of the BTE is desirable. Angular and space discretization can be implemented by combining the discrete ordinates method (DOM) and the finite volume method (FVM) ~\cite{Hunter2015,Samian-2014,Hamian2015,Vallabhaneni2017}. However, as the advection term in the FVM is not reproduced exactly, numerical smearing and angular false scattering is introduced \cite{Hunter2015}. This effect is particularly strong if, as frequently done, a simple upwind discretization scheme is used. Recently, the finite-volume based discrete unified gas kinetic scheme (DUGKS) for phonon transport has been proposed \cite{GUODUGKS2016}. DUGKS makes use of the method of characteristics instead of the upwind scheme to calculate the convective term.

The LBM is a fully Lagrangian method where advection is performed exactly along the characteristics (as the grid exactly matches the advection) \cite{Guo1999}. Despite this favorable property, two severe issues hamper the use of the standard LBM for simulations in the high Knudsen number regime. Firstly, the low number of available propagation directions on the grid introduces a strong ray effect (e.g. 8 for the standard D2Q8 lattice in 2D, Fig. \ref{fig:D2Q6-D2Q8}a) \cite{nabovati2011}. Secondly, regular square grids suffer from non-isotropic speed of propagation.  For instance, in the D2Q8 lattice the propagation speed along the diagonal directions (directions 5-8, Fig. \ref{fig:D2Q6-D2Q8}a) is a factor of $\sqrt{2}$ faster than  the one along the axial directions (directions 1-4, Fig. \ref{fig:D2Q6-D2Q8}a). In the framework of the LBM, the 2D grid with the highest number of propagation directions (six directions) with isotropic velocity is the hexagonal D2Q6 (Fig. \ref{fig:D2Q6-D2Q8}b). The underlying idea of the D2Q6 is that all distances to the next neighbors are equal from each grid point. Extension of this idea to 3D is in principle possible, because Platonic solids fulfill the requirements. Still, the variants are strongly limited because only one Platonic solid, the cube, leads to a completely space-filling tiling of the 3D space  \cite{torquato2009}. Therefore, the grid with the highest number of propagation directions with isotropic velocity in 3D is the D3Q6. Both the D2Q6 and the D3Q6 are insufficient to overcome the ray effect problem and, hence, are unsuitable for ballistic transport \cite{nabovati2011}.
Other approaches have been proposed to increase the number of propagation directions. For instance,  it has been suggested to use enlarged stencils, i.e. high order velocity sets, which propagate among next neighbor lattice points and beyond \cite{Thouy2008}. Higher-order lattices have also been obtained by discretizing the Boltzmann equation using the roots of Hermite polynomials ~\cite{Shan1998}. The limitation of this method is that the corresponding discrete velocities cannot be incorporated into a regular space-filling lattice. This leads to the loss of the main advantage of the LBM, which is aimed to be used in this work: the exact space discretization of the advection term \cite{Frapolli2014}. 

In this work, to solve the velocity anisotropy related issue,  we propose the time adaptive scheme lattice Boltzmann method (TAS-LBM). The TAS-LBM allows to describe an isotropic speed of propagation on cartesian grids. It can also be extended beyond the isotropic speed of propagation to describe any angle-dependent velocity. Finally, and most importantly,  we solve the ray effect issue of the LBM by introducing an arbitrary high (in 2D multiples of 8) number of grid propagation directions, following the approach of Thouy et al. (i.e. propagating among next neighbour lattice points and beyond)~\cite{Thouy2008}. However, to reduce  the computational cost and complexity  that results from increasing the number of propagation directions \cite{Thouy2008,Pietro2011}, we propose the worm-LBM. This method allows to approximate arbitrary directions by alternating in time the basic directions provided by the next neighbor schemes. For convenience, an algorithm allowing schemes with an arbitrary high number of directions (multiples of 8) is described.

The range of applications for the new worm-LBM, in principle, extends to all transport problems with dominating hyperbolic character. These problems show solutions along characteristics  $u(t,\vec{x})=u_0(\vec{x}-\vec{v}t)$, where $u_0$ is the initial condition, $\vec{x}$ is space, $t$ is time, and $\vec{v}$ is the velocity. The improvement in the description of the discontinuity properties in hyperbolic problems is of interest for a wide variety of fields of research  (see e.g. LeVeque \cite{LeVeque2002} for a comprehensive overview). Here, as it is of high practical relevance, we introduce the new algorithm for solving the BTE, which is a representative example of a first-order transport equation. Note that the BTE is parabolic or hyperbolic depending on the scattering operator. For the Bhatnagar-Gross-Krook operator (BGK),   known as the relaxation time approximation \cite{Bhatnagar1954}, the BTE shows a parabolic character and becomes hyperbolic in the limit of weak scattering (i.e. ballistic regime). 

In the following, the proposed methods will be explained in detail, and the numerical accuracy will be verified by comparison with analytical solutions in the Fourier and the ballistic limit, using different initial and boundary conditions.  Moreover, a comparison to Monte Carlo simulations will be provided for one of the benchmark tests.

\section{Models and numerical methods}
\label{sec:methods}
\subsection{Peierls-Boltzmann phonon transport equation}
In semiconductors and dielectric materials the phonon contribution to the thermal transport dominates, the electron contribution is negligible, and the heat transport problem reduces to solve the phonon Peierls-BTE. Accordingly:
\begin{equation}
    \frac{\partial\,f}{\partial\,t}\,+\,\vec{v}_\text{g}\cdot\nabla\,f =\left(\frac{\partial\,f}{\partial\,t}\right)_\text{scattering}
    \label{eq:bte}
\end{equation}
where $f(\vec{x},t,\vec{k}, p)$ is the phonon distribution function of a phonon state ($\vec{k},p$)  at position $\vec{x}$ and   time $t$. Here, $\vec{k}$ accounts for the wave-vector, and $p$ for the polarization; while  $\vec{v}_\text{g}$ is the group velocity related to the phonon frequency through the dispersion relation $\vec{v}_{\text{g}}(\vec{k},p)=\nabla_{\vec{k}}\,\omega(\vec{k},p)$. Adopting polar coordinates,  the state  ($\vec{k},p$) can be written as $(k, \hat{s}, p)$,   where $k=|\vec{k}|$ is the wave number and  $\hat{s}=(\theta,\phi)$ provides the direction ($\theta$ and $\phi$ are the polar and azimutal angle, respectively). Thus, using the frequency instead of the wave number, the phonon distribution function is transformed to: $f(\vec{x},t,\vec{k}, p)\rightarrow f(\vec{x},t,\hat{s},\omega, p)$. Note that all variables in Eq. \ref{eq:bte}  are space and time dependent. In the following, for the sake of clarity,  we omit $(\vec{x},t)$ and use only $(\hat{s},p)$ as indexes [e.g. $f_{\hat{s},p}(\omega)$ = $f(\vec{x},t,\hat{s},\omega, p)$].

The second term on the left-hand side of Eq. \ref{eq:bte}  describes the advection of phonons, while the term on the right-hand side denotes the phonon scattering. The latter, in the  relaxation time approximation (RTA), can be written as:
\begin{equation}
    \left(\frac{\partial\,f_{\hat{s},p}(\omega)}{\partial\,t}\right)_\text{scattering}  = \frac{f_\text{BE}^\text{eq}(\omega, T)-f_{\hat{s},p}(\omega)}{\tau_{\hat{s},p}(\omega)}.
\end{equation}
The RTA assumes that a non-equilibrium phonon distribution function $f_{\hat{s},p}(\omega)$ 
relaxes within a time $\tau_{\hat{s},p}(\omega)$  to the local equilibrium Bose-Einstein distribution: 
\begin{equation}
f_\text{BE}^\text{eq}(\omega,T) = \frac{1}{\exp[{\hbar\,\omega/k_\text{B}T}]-1},
\label{eq:Bose-Einstein}
\end{equation}
where  $T$ is the equilibrium temperature,  $\hbar$ is the Planck constant divided by $2\pi$, and $k_\text{B}$ is the Boltzmann constant. 
 The individual phonon scattering contributions to the relaxation time (i.e. phonon-phonon, phonon-electron, phonon-defect scattering) sum up according to the Matthiessen rule $1/\tau=\sum_i{1/\tau_{i}}$. 
 
 Equation \ref{eq:bte} can be written in terms of the direction and polarization dependent energy density:
\begin{eqnarray}
\label{eq:bte-density} 
\frac{\partial e_{\hat{s},p}(\omega)}{\partial t} +\vec{v}_{\hat{s},p}(\omega)\cdot\,\nabla\,e_{\hat{s},p}(\omega) &&\\ =\frac{e^\text{eq}(\omega,T)-e_{\hat{s},p}(\omega)}{\tau_{\hat{s},p}(\omega)},&&\nonumber
\end{eqnarray}
where
\begin{equation}
e_{\hat{s},p}(\omega)=\int \hbar\,\omega\,f_{\hat{s},p}(\omega)\,D_{\hat{s},p}(\omega)d\omega.
\end{equation}
Here, $D_{\hat{s},p}(\omega)$ is  the direction and polarization dependent phonon density of states per unit volume; and the equilibrium energy density is given by 
\begin{equation}
e^{eq}(\omega,T) = \sum_{p}\int_{\Omega}\int \hbar\,\omega\,f_\text{BE}^\text{eq}(\omega,T)\,D_{\hat{s},p}(\omega)\,d\omega d\Omega,
\label{eq:eq-energy}
\end{equation} 
where $\Omega=\sin\theta \,d\theta \,d\phi$ is the solid angle. It is also common to write the BTE in terms of the  phonon intensity, which is related to the directional energy density by $I=v_g\cdot e/4\,\pi$.

Note that the advection term allows for an arbitrary number of propagation directions ($\hat{s}$), which are coupled by the scattering term. This renders  Eq. \ref{eq:bte-density}  into  an infinite set of coupled differential equations. Therefore,  numerical schemes using direct discretization have to include, aside of time and space, angular discretization.

In this work, for simplicity, all the methods are introduced in the framework of the gray approximation. Accordingly,  the BTE is solved to predict the distribution function of only one representative phonon mode with average properties (i.e. all phonons
have the same polarization, uniform group velocity, and relaxation time). Thus, the density of states can be written as $D(\omega) = 3 N \delta (\omega - \omega_0)$, where  $N$ is the number of atoms,  and $\omega_0$ the phonon frequency. Replacing the density of states into Eq. \ref{eq:eq-energy}, the equilibrium energy density in the gray model can be write as:
\begin{equation}
e^{eq}(T) = \frac{3 N}{V} \,\hbar\,\omega_0 \,f_\text{BE}^\text{eq}(\omega_0,T),
\label{eq:gray-energy-density}
\end{equation}
where $V$ is the unit-cell volume. Nonetheless, all the  methods presented here can be straight forward extended to account for an  arbitrary phonon band structure. This is known as the dispersion LBM ~\cite{nabovati2011,Sellan2010}. Alternatively, the "multigray" approach can be also adopted. This method  solves multiple uncoupled gray BTEs in parallel, and uses the superposition principle to obtain the final result \cite{Romano2019}.
\subsection{The phonon lattice Boltzmann method}
\label{sub:LBM}

\begin{figure}
    \centering
    \includegraphics[scale=0.53]{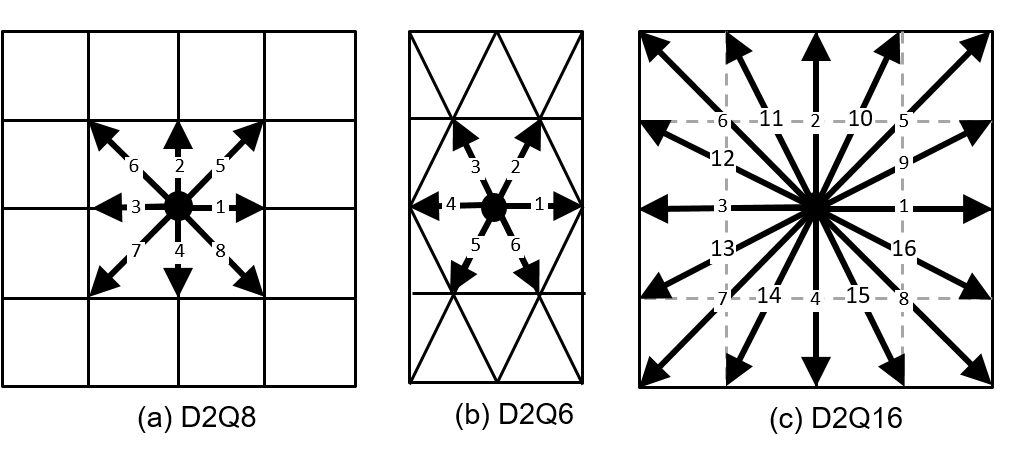}
    \caption{Standard D2Q6, D2Q8 and D2Q16 grids in 2D, with Q= 6, 8 and 16 propagation directions, respectively. }
    \label{fig:D2Q6-D2Q8}
\end{figure}

Discretizing the first derivatives in Eq. \ref{eq:bte-density} with respect to time and space:
\begin{eqnarray*}
    \frac{\partial e_{\hat{s}}}{\partial t}&\approx&\frac{e_{\hat{s}}(\vec{x},t+\Delta t)-e_{\hat{s}}(\vec{x},t)}{\Delta t}\\ 
    \frac{\partial e_{\hat{s}}}{\partial \vec{x}}&\approx&\frac{e_{\hat{s}}(\vec{x}+\Delta \vec{x},t+\Delta t)-e_{\hat{s}}(\vec{x},t+\Delta t)}{\Delta \vec{x}},
\end{eqnarray*}
and introducing angular discretization ($i$) to account for the phonon propagation in a discrete grid direction $\hat{s}_i$ (see Fig. \ref{fig:D2Q6-D2Q8}), we define the discrete directional energy density (DDED) $e_i$. Thus, Eq. \ref{eq:bte-density} can be then written as:
\begin{eqnarray}
        \frac{e_i(\vec{x},t+\Delta t)-e_i(\vec{x},t)}{\Delta t} &+& \notag\\ 
        \vec{v}_{\text{g}\,i}\frac{e_i(\vec{x}+\Delta \vec{x}_i,t+\Delta t)-e_i(\vec{x},t+\Delta t)}{\Delta \vec{x}}&&\notag \\=\;\frac{e_i^\text{eq}(\vec{x},t)-e_i(\vec{x},t)}{\tau}.&&
        \label{eq:discrete-BTE}
\end{eqnarray}   
The energy density has to be conserved by the the collision operator, i.e. the sum of both sides of Eq. \ref{eq:discrete-BTE}, over all grid points has to give a constant value over time. This condition is fulfilled by local equilibrium energy densities calculated as: 
\begin{equation}
    e^{eq}(\vec{x},t)=\sum_{i}^{Q} e_i(\vec{x},t),
    \label{eq:lbm-energy-density}
\end{equation}
where the sum runs over all directions $Q$ on the grid.
After scattering, the energy is redistributed according to:
\begin{equation}
    e_i^{eq}= \sum_{i}^{Q} w_i e^{eq}(\vec{x},t),
    \label{eq:lbm-energy-weighting}
\end{equation}
where $w_i$ is an angular weighting factor, with $\sum_i w_i = 1$, accounting for the proportion of solid angle represented by a given direction (see explanation in Sec. \ref{sec:ray-effect}). In simple cases in which the angular space is divided into uniform intervals (e.g. D2Q8, Fig. \ref{fig:D2Q6-D2Q8}a), and assuming isotropic scattering:  $e_i^{eq}=e^{eq}/Q$. 

In the LBM approach, the space domain is discretized by defining a grid of points separated by a lattice vector $\Delta \vec{x}_i = \vec{c}_i \Delta t$, where  $\vec{c}_i$ is a discrete lattice velocity vector. Thus, the energy densities and derived macroscopic quantities are defined at discrete positions on a grid (e.g. in 2D: $\vec{x}=(x_j,y_k)$, with $j,k\in\mathbb{Z}$), and discrete times $t=t_n$  ($n\in\mathbb{N}$).  Rearranging Eq. \ref{eq:discrete-BTE}, the LBM algorithm can be written as:
\begin{equation}
    e_i(\vec{x}+ \vec{c}_i \Delta t,t+\Delta t) = (1-W)\,e_i(\vec{x},t)+W\, e_i^\text{eq}(\vec{x},t).
    \label{eq:lbm-energy}
\end{equation}
Here, 
\begin{eqnarray}
    \label{eq:W-Kn}
    &W \;=\;\frac{\Delta t}{\tau} \;=\; \frac{\Delta x}{v_\text{g}}\frac{1}{\tau} \;=\;\frac{1}{N}\frac{1}{\text{Kn}}&\\
    &(\Delta x=\frac{L}{N},\; \text{Kn}=\frac{\Lambda}{L},\;\Lambda =v_\text{g}\tau),&\nonumber 
    \end{eqnarray}
is a weighting factor setting the discrete directional energy density (DDED) (first term RHS of Eq. \ref{eq:lbm-energy}), and the equilibrium energy density (EED) (second term RHS of Eq. \ref{eq:lbm-energy}) contributions. $L$ and $N$ are, respectively, the length of the domain and the number of grid points in which it is divided, $\text{Kn}$ denotes the Knudsen number, and $\lambda$ is the phonon mean free path (MFP). Note that the weight of the scattered term EED decreases with the grid refinement. As the MFP must not change with a finer grid, the  DDED has to propagate more grid points per physical length. Hence,  less scatter takes place while propagating one lattice spacing. Equation \ref{eq:lbm-energy} is the core of the LBM method, and it is common to most LBM algorithms.

The local "pseudo" temperature, or also called effective temperature, at a given time ($t$), and lattice site ($\vec{x}$) can be calculated by inserting the local equilibrium energy density computed in Eq. \ref{eq:lbm-energy-density} into Eq. \ref{eq:gray-energy-density}:
\begin{equation}
    T(\vec{x},t,e^{eq})= \frac{\hbar\,\omega_0}{k_\text{B}\,\ln(\frac{3\,N\,\hbar\,\omega_0}{V\,e^\text{eq}}+1)};
    \label{eq:pseudo-T}
\end{equation}
while the heat flux vector can be calculated according to:
\begin{equation}
\vec{q}(\vec{x},t) = v_\text{g} \sum_{i}^{Q} e_i \frac{\vec{c_i}}{|\vec{c_i}|}  .
    \label{eq:q_corr}
\end{equation}

Mind that the effective temperature (Eq. \ref{eq:pseudo-T}) provides a comparison between the energy density of a non-equilibrium system  and one for which temperature is well defined (i.e. a system at local equilibrium).

\subsection{Solving the lattice anisotropy issue: the time adaptive scheme}
\label{sec:TAS}

In the gray approximation, as well as in the isotropic formulation of the dispersive-LBM, a specific phonon mode travels with the same speed independent of the propagation direction. As discussed in the introduction, Nabovati et.al.  \cite{nabovati2011} have shown that the two dimensional hexagonal scheme D2Q6 (Fig. \ref{fig:D2Q6-D2Q8}b) provides the grid with the highest number of isotropic lattice vectors. Thus, this grid has the  highest number of propagation directions with isotropic speed (six directions). Although this grid is suitable to describe the diffusive regime~\cite{nabovati2011}, it poorly describes the ballistic one. The D2Q8, alternatively, resolves two more propagation directions, allows easier handling of the boundary conditions, and provides a natural extension to 3D \cite{GUO20161}. However, in this case, the lattice velocity along the diagonal directions is $\sqrt{2}$ times faster than along the axial ones, resulting in an unrealistic phonon propagation ~\cite{nabovati2011}.  In the diffusive limit (i.e. Kn$\,\ll$ 1),  this velocity error translates into a "grid MFP" $\Lambda^\text{grid}_{i}=\vec{c}_{i}\tau$ that is a factor of $\sqrt{2}$ longer in the diagonal directions. In this case, a simple solution is to adjust the corresponding $\tau_{i}$ by a factor of $2/(1+\sqrt{2})$.
This correction, however, does not solve the velocity issue in the ballistic limit.

\begin{figure}
    \centering
    \includegraphics[scale=0.9]{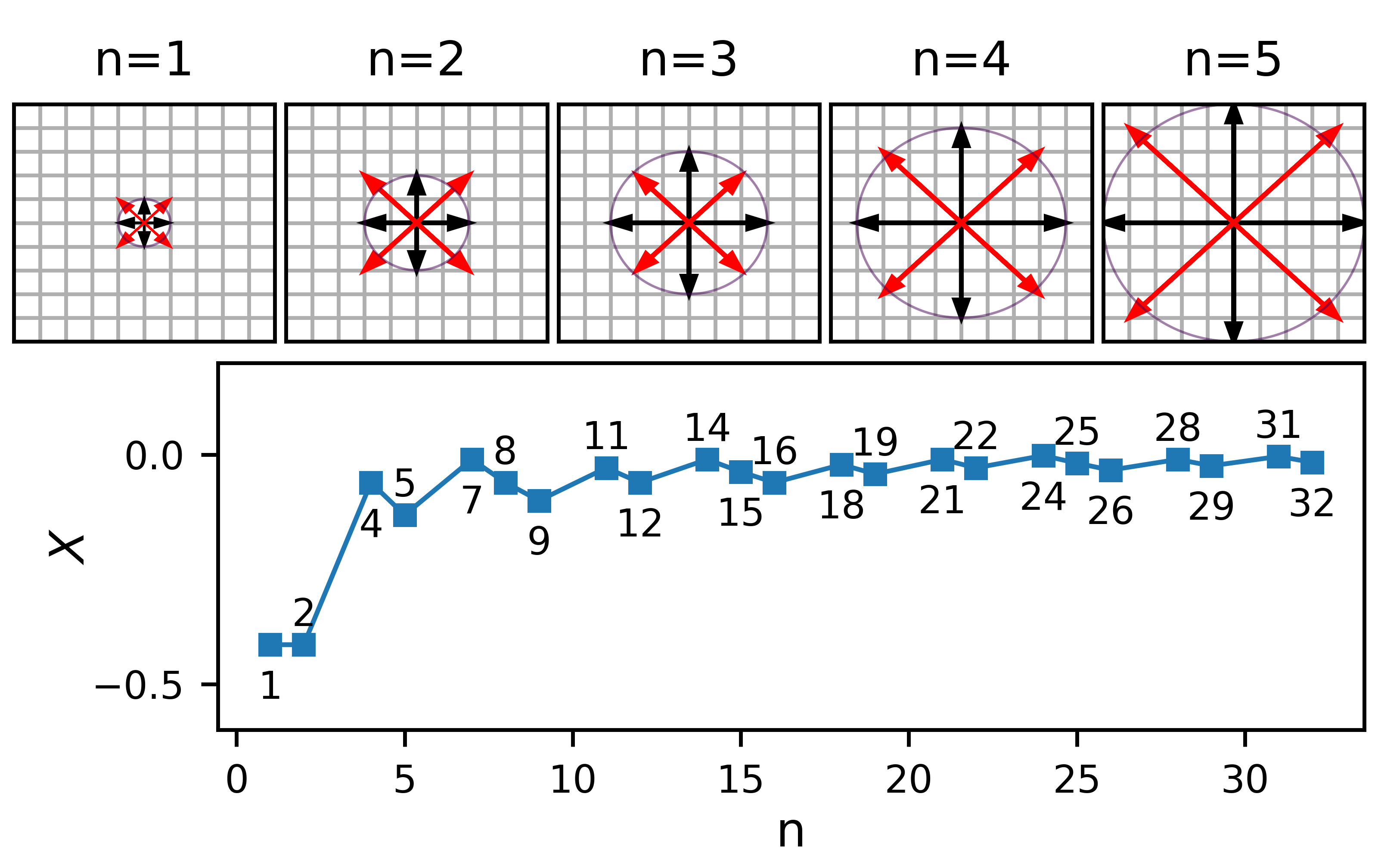}
    \caption{Upper panel: Hopping evolution during the first five time steps along the D2Q8 grid. The arrows indicate the propagation directions and the accumulated distance at each time step $n$. Lower panel: Relative distance,  $X=(\Delta x_{\mathrm{diag}}-\Delta x_\mathrm{ax})/\Delta x_\mathrm{ax}$, propagated along the diagonal direction ($x_\text{diag}$) with respect to the axial one ($x_\text{ax}$) as a function of time $n$.  The numbers within the graphic indicate the time steps $n$ at which diagonal hopping takes place.}
    \label{fig:hopping}
\end{figure}

To overcome the problem on regular grids (e.g. the D2Q8) in the ballistic regime, we propose the  time adaptive scheme lattice Boltzmann method (TAS-LBM). This method imposes a time condition for the propagation of the DDED's contribution to the LBM, such that phonons are allowed to propagate along the diagonal direction only if they remain within a circle of radius $r_\text{ax}$. This radius is given by the distance propagated along the axial directions $x_\text{ax}$ plus one lattice spacing $\Delta x_i$ (see Fig. \ref{fig:hopping}). Otherwise, the propagation along the diagonal direction is paused until the condition is fulfilled.  The scattering and propagation of the EED part of the LBM is, nonetheless, executed for those directions for which their DDED propagation is paused along the diagonal. In a nutshell, the algorithm for the TAS-LBM can be written as:
\begin{align}
&\text{if } x_\text{cum} \leq x_\text{ax}+\Delta x_i\text{:}\quad \text{(propagating all)}
\label{eq:algorithm}\\
&\; e_i(x+\Delta x_i,t+\Delta t)  =  (1-W)\,e_i(x,t)+W\, e_i^\text{eq}(x,t) \nonumber\\
&\text{if }x_\text{cum} > x_\text{ax}+\Delta x_i\text{:}\quad \text{(pausing DDEDs)}\nonumber\\
&\;e_i(x+\Delta x_i,t+\Delta t) =  (1-W)e_i(x+\Delta x_i,t)+W e_i^\text{eq}(x,t),\nonumber
\end{align}
where $x_\text{cum}$ and $x_\text{ax}$ are, respectively, the cumulative  distances propagated along the direction $i$, and  the axial directions at a certain time step. For the axial directions,  the first condition in Eq. \ref{eq:algorithm} is always fulfilled. The upper panel of Fig. \ref{fig:hopping} shows the hopping evolution during the first five time steps along the D2Q8 grid. As it can be seen, diagonal hopping takes place at time steps $n=1,2,4,5$, while at $n=3$ it is paused. As the simulation evolves in time, the  distance propagated along the diagonal direction converges toward the distance travelled along the axial axis (lower panel Fig. \ref{fig:hopping}), which over time results in  an isotropic speed of propagation.  
In the limit of $W=1$, only the evolution of the EED term is contributing, and the algorithm reduces to the standard D2Q8.  On the other hand, if $W$ is close to zero, i.e.  when the DDED contribution is dominating, the velocity problem of the standard D2Q8 is repaired by the TAS-D2Q8. Note that values of $W$ lower than one can be cheaply fulfilled in the limit of high Knudsen numbers (ballistic regime),  even with a relatively coarse grid ($W=1/(N \text{Kn})$ see Eq.\ref{eq:W-Kn}).  For instance, $W=0.1$ can be obtained for $\text{Kn}=1$ using a coarse grid with $N=10$ grid points. 

Special care must be taken when comparing 3D  results (analytical or from other numerical methods) to 2D simulations. On the one hand, it should be taken into account that phonons can also move ballistically out of plane in 3D, leading to the following relationship between velocities \cite{nabovati2011} :
\begin{equation}
    v_\text{g}^{\text{2D, effective}} = \frac{1}{\pi} \int_{0}^{\pi} v_\text{g}^{\text{3D}} \sin{\theta} \,d\theta = \frac{2}{\pi} v_\text{g}^{\text{3D}}.
    \label{eq:proj3D2D-ballistic}
\end{equation}
On the other hand, there is a 2/3 factor between the diffusivity in 2D and 3D, as can be seen from comparing their corresponding mean square displacements ($\langle \vec{x}^2\rangle$):
\begin{align}
    2D:\;\;\;\; \langle x^2+y^2 \rangle &= 2 \langle x^2 \rangle = 2 D_2 t\nonumber \\
    3D: \langle x^2+y^2+z^2  \rangle &= 3 \langle x^2 \rangle = 2 D_3\nonumber t\\
    \rightarrow D_2 &= 2/3 D_3.
    \label{eq:proj3D2D-diffusive}
\end{align}
Here $D_2$  and $D_3$ are the diffusivity in 2D and 3D, respectively. This effect, as proposed by Guo and Wang \cite{GUO20161}, can be accounted for by rescaling the lattice velocity with respect to the group velocity (e.g. $c=2/3 v_{g}$ for the D2Q8 lattice). They derived this scaling factor by combining the Chapman-Enskog expansion and the Fourier heat equation \cite{GUO20161}, which in practice consists in adopting the diffusive Fourier limit. The problem of this approach is that it leads to a wrong phonon propagation speed in the transient ballistic regime, where the phonon "wave front" would arrive too late (e.g.  $2/3$ slower than the phonon velocity defined by the group velocity along axial direction, and $2/3\,\sqrt{2}$ slower in the diagonal one for the D2Q8 grid). 
In this work, to account for this effect we follow a rather different approach.   After setting the grid spacing by computing $W$ according to the desired Knudsen number (see Eq. \ref{eq:W-Kn}), a new  corrected weighting factor $W_{corr}$ is used in the LBM algorithm  (Eq. \ref{eq:algorithm}):  
\begin{equation}
    W_{corr} = \frac{3}{2}[(1-W) W + W (W \Lambda_{corr})],
    \label{eq:W-correction}
\end{equation}
where the $3/2$  factor accounts for diffusivity difference in 2D and 3D (this factor should be used only in those cases in which a 2D simulation is used to emulate a 3D behavior), and 
\begin{equation}
    \Lambda_{corr} = \frac{2}{1+\sqrt{2}}
    \label{eq:W-correction-factor}
\end{equation}
accounts for the velocity issue in the diffusive limit for the D2Q8 grid (i.e. different velocities along the axial and diagonal directions).  As the correction is done after setting the grid spacing, the velocities are not modified ($\Delta x = v_\text{g} \Delta t $). Thus, the correction is applied to the  MFPs (by modifying the relaxation times) and not to the velocities.  Eq. \ref{eq:W-correction} results as a consequence of mixing the advection of the ballistic non-equilibrium term (first term RHS in Eq. \ref{eq:lbm-energy}),  and the transport of the scattered EED term (second term RHS in Eq. \ref{eq:lbm-energy}). The latter  has to be corrected since it always propagates as in the standard D2Q8. Thus, in the limit of $W \rightarrow 0$,  $W_{corr} = 3/2$, and in the limit of $W \rightarrow 1$,  $W_{corr} = 3/2\,\Lambda_{corr}$.
\begin{figure}
    \centering
    \includegraphics[scale=0.5]{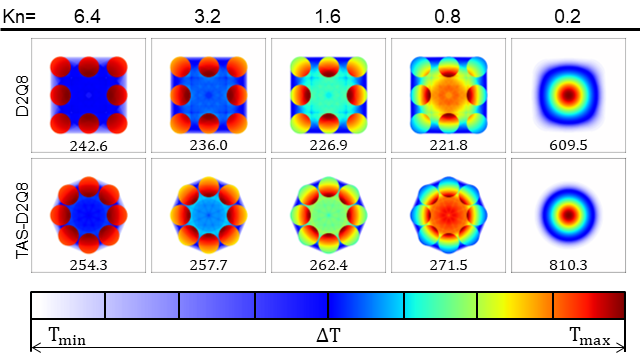}
    \caption{
     Heat propagation from a circular high-temperature region into a  colder infinite domain for five different Knudsen numbers, as obtained with the standard LBM (D2Q8) and the new TAS-LBM (TAS-D2Q8). The initial temperature of the domain, except for a central region of radius $1/11$ of the showed domain length, was set to 299 K. The central region was initialized at 301 K.  The total number of grid points was N = 512, while the Knudsen number took values of 6.4, 3.2, 1.6, 0.8 and 0.2. The color map starts a the minimum Temperature $T_{\text{min}} = 299$ K (white), and is re-scaled to the highest temperature occurring in the simulation (dark red). The latter is given by the temperature difference $\Delta T = T_{\text{max}}-T_{\text{min}}$ in mK as shown below each sub-figure.}   
    
    \label{fig:LBM-TAS}
\end{figure}

The performance of the standard (without MFP correction) LBM (D2Q8) and the TAS-D2Q8 is compared in Fig. \ref{fig:LBM-TAS}.  The heat dissipation from a circular high-temperature region into a colder infinite domain is shown. Five different Knudsen numbers encompassing regimes from the ballistic to the diffusive one are considered.  The differences between LBM and TAS-LBM are evident in the ballistic regime.  Heat propagates faster along the diagonal directions within the D2Q8-LBM scheme, while it propagates isotropically within TAS-LBM.  In the diffusive regime, the non-isotropic speed of propagation effect is more subtle. It can be seen, however, that the different contours for the D2Q8-LBM case are more extended than those obtained with the TAS-LBM, as heat is propagated faster.  Thus, by the use of a simple time-scheme approach, the TAS-LBM  allows to describe a  circular propagation on a square grid, making the method particularly suitable for the gray approximation and the isotropic dispersion case.  The TAS-LBM, nonetheless, is not restricted to isotropic conditions and can be used to impose any prescribed angle-dependent velocity distribution. 

\subsection{Solving the ray effect issue, first variant: next-next neighbor advection}
\label{sec:ray-effect}

In the previous section, we have shown that the TAS-LBM allows isotropic speed of propagation for the advection term on a D2Q8 grid. Nevertheless,  as a consequence of the finite number of discrete propagation directions, the angular-dependent temperature profiles feature unrealistic bumps and oscillations as the Knudsen number increases (see temperature profiles in Fig. \ref{fig:LBM-TAS}).  This problem is known as the ray effect, and it is intrinsic of direct discretization methods~\cite{nabovati2011}. 

To make the TAS-LBM also accurate in the ballistic regime, it is  desirable to have more directions along which the phonons can travel on the grid.  The number of directions on a square grid can be augmented by increasing the number of grid points used for advection. For instance, one can resolve eight additional propagation directions for the phonons by advecting to the next-next neighbours in the D2Q8 grid  \cite{Thouy2008} (D2Q16, Fig. \ref{fig:D2Q6-D2Q8}). In total, there would be 16 directions encompassing three different lattice velocities along the axial (1-4), diagonal (5-8) and, intermediate (9-16) directions. In this case, the TAS-LBM will enable propagation along the diagonal and intermediate directions,  as long as the propagated distance is smaller than or equal to the distance traveled along the axial direction at a given time step (i.e. the distance hopped along directions 1-4, Fig. \ref{fig:D2Q6-D2Q8}).  Thus, the TAS-LBM scheme can be easily generalized to enforce a circular propagation for any D2Q[$M\times$8] grid, $M$  being an integer. These grids would allow to resolve $M\times$8 propagation directions, improving the ray effect problem, and providing a more accurate description of phonon transport in the ballistic regime. It is important to remark again that the TAS-LBM goes beyond the isotropic condition, and that it can be used to describe any angular distribution of velocities.

\begin{figure}
    \centering
    \includegraphics[scale=0.6]{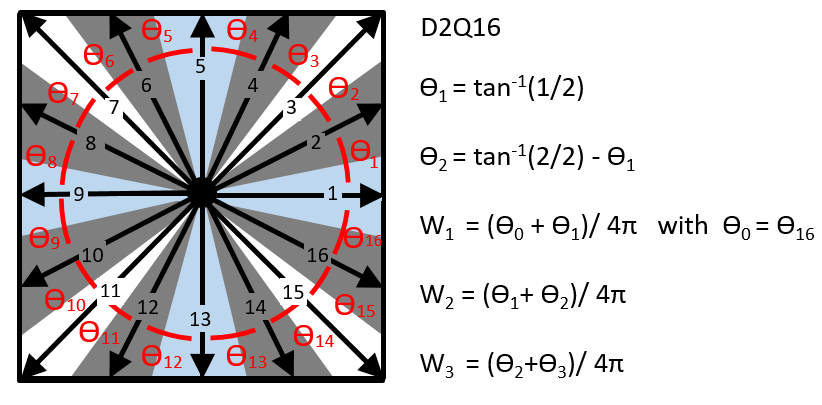}
    \caption{ Example of weight distribution for the D2Q16 grid. Only the weights for the first three directions are listed, the remaining ones can be straightforwardly calculated  by using Eq. \ref{eq:weight}. Note that, $\theta_1=\theta_4=\theta_5=\theta_8=\theta_9=\theta_{12}=\theta_{13}=\theta_{16}$, and $\theta_2=\theta_3=\theta_6=\theta_7=\theta_{10}=\theta_{11}=\theta_{14}=\theta_{15}$.    }
    \label{fig:angle}
\end{figure}

For D2Q[$M\times$8] grids with $M>1$, the angles between adjacent propagation directions get smaller towards the diagonals. To compensate this effect and to take into account the angular interval that should be represented by one direction, an angular weighting function should be introduced. In practice, for a given scheme D2Q[$M\times$8] there will be a set of $M\times$8 angles,
\begin{equation}
\theta_j=\tan^{-1}(j/M), \;\; (j=1,...,M\times8),  
\end{equation}
that define $M\times$8 weight factors:
 \begin{equation}
    w_{j} = \frac{\theta_{j-1} + \theta_{j}}{4\,\pi} \;\; (1<j<M\times8), 
    \label{eq:weight}
     \end{equation}
where angular periodic boundary conditions are considered (i.e. $\theta_{0}=\theta_{M\times8}$). An example for D2Q16 grid is provided in Fig. \ref{fig:angle}. 
These weight factors are  used for the initial and boundary conditions, and for distributing the post-scattering equilibrium energy densities to the different lattice directions (see Eq. \ref{eq:lbm-energy-weighting}). Moreover, the grid-MFP correction factor (Eq. \ref{eq:W-correction-factor}) has to be generalized for schemes with arbitrary number of $Q$ directions:
\begin{equation}
    \Lambda_{corr} = \frac{Q}{\sum_i^Q c_i} .
    \label{eq:W-correction-factor-gereralized}
\end{equation}

\subsection{Solving the ray effect issue, final variant: the worm-LBM}
\begin{figure}[h]
    \centering
    \includegraphics[scale=0.5]{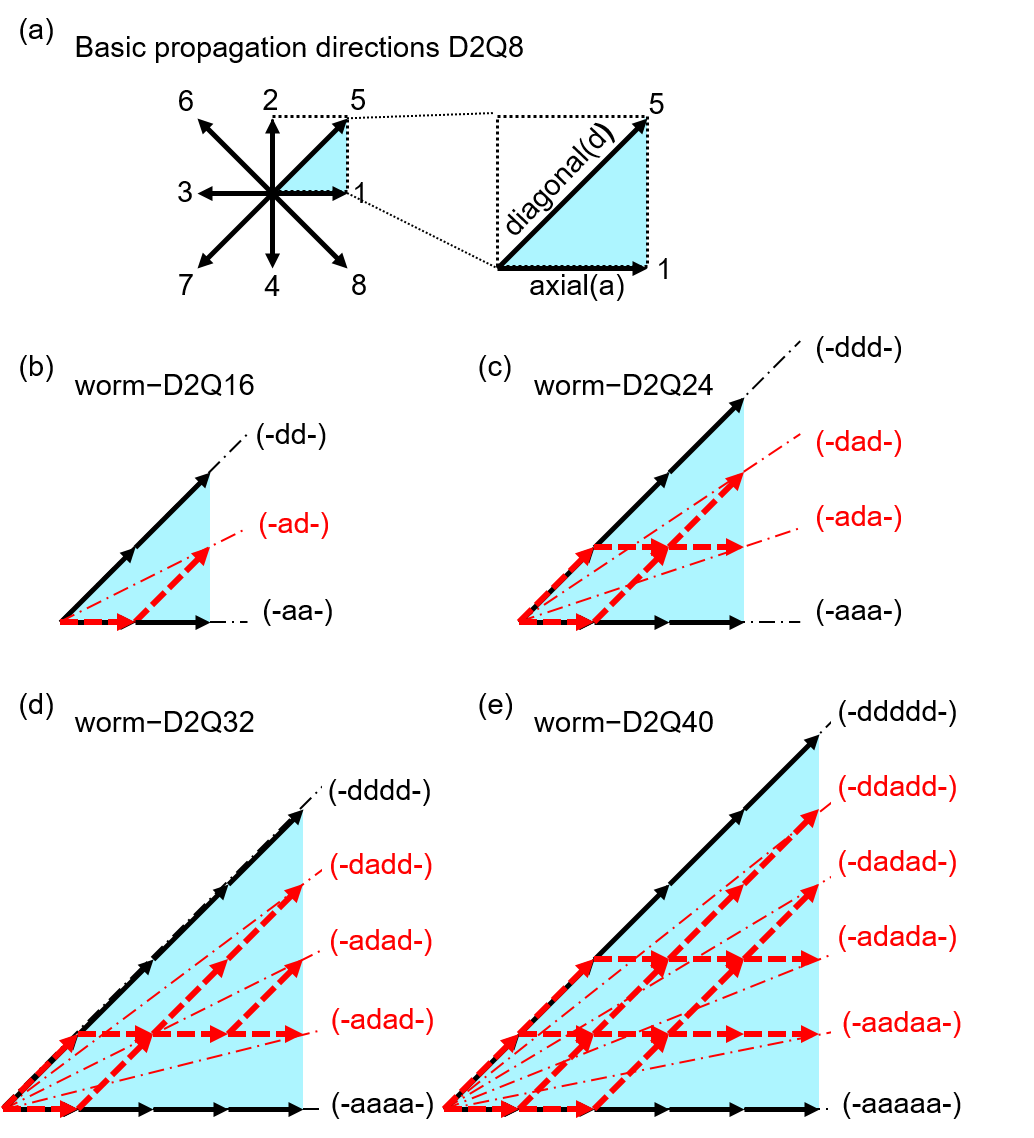}
    \caption{The worm-LBM approach. Propagation on any grid D2Q[$M\times8$] is constructed from the basic directions described within the (a) D2Q8 grid. Worm-path examples:  (b) worm-D2Q16, (c) worm-D2Q24, (d) worm-D2Q32 and  (e) worm-D2Q40. The additionally introduced propagation directions along the D2Q[$M\times8$] are described by the dash-dotted lines, while the worm-path composed of axial, a, and diagonal, d, lattice directions are  denoted by dashed arrows (listed in round parenthesis).  Note that the directions that are not shown in the first quadrant, and on the other three quadrants, are symmetrically equivalent to those depicted.}
    \label{fig:CLBM-worm}
\end{figure}

The TAS-LBM provides a reliable approach to correctly describe both the ballistic and the diffusive regimes of phonon transport. Nonetheless, schemes using the next-next neighbor advection (and so forth) have a reduced resolution in space for representing initial conditions, boundary conditions, and inhomogeneities in the material parameters.  For instance, while the standard D2Q8 next neighbor advection scheme allows 1$\times$1 structure sizes, the D2Q16 next-next neighbor scheme requires 2x2 structure sizes (i.e. four times more grid points compared to the D2Q8); the D2Q24 next-next-next neighbor scheme requires 3x3 structure sizes (i.e. nine times more grid points), and so forth. Thus, the number of grid points has to increase with the number of directions to keep the space resolution, which increases the computational cost. An alternative to circumvent this limitation is to encode the spatial propagation in a square grid D2Q[M × 8] (with M > 1) into a time adaptive propagation on the D2Q8 scheme (i.e. propagating in time using the directions defined within the D2Q8 next neighbor scheme as building blocks). It can be shown that any propagation direction of a  D2Q[$M\times8$] scheme can be decomposed into the basic directions described  within a standard D2Q8 stencil: i.e. using the axial directions 1-4 and the diagonal directions 5-8, as shown in Fig. \ref{fig:CLBM-worm}(a). In practice, the diagonal and axial directions of the D2Q8 scheme are repeated in a optimized sequence (worm-path), such that the trajectory in a desired direction is followed as close as possible. The worm-path is repeated every $M$ time steps (worm-sequence time steps), and can be defined at the beginning of the simulation using the following algorithm.  Being $l^{\hat{s}}_{\text{Q}[M\times8]}$ the total distance propagated by the advection term after $M$ time-steps in a direction ($\hat{s}$) within the D2Q[$M\times8$] scheme, and $P_{\hat{i}}$ the nearest neighbour grid point in a direction $\hat{i}$ from the point where advection is taking place,  the worm-path finding algorithm reads as:
\begin{align}
\label{eq:worm-paths}
&\text{for} \,\,\hat{s} = 1, \, Q = M\times8: \\
& \text{\footnotesize (loop over the  D2Q[$M\times$8] directions)}\nonumber\\
&\qquad\text{for} \,\,n = 1, \, M: \nonumber\\
&\qquad \text{\footnotesize  (loop over the worm-sequence time steps)}\nonumber\\
& \qquad\qquad \delta = 1 \nonumber\\
& \qquad\qquad \text{for} \,\,\hat{i} = 1,\, 8: \nonumber\\
& \qquad\qquad\text{\footnotesize (loop over the D2Q8 directions)}\nonumber\\
& \qquad\qquad\qquad \delta_{i-1} = \delta \nonumber\\
& \qquad\qquad\qquad \delta_i = \text{distance}\big[l_{\text{Q}[M\times8]} ^{\hat{s}}\,\frac{8\,n}{M}, P_{\hat{i}} \big] \nonumber\\
& \qquad\qquad\qquad \text{if} \,\, \delta_i \leq \delta_{i-1}\,:\nonumber\\
& \qquad\qquad\qquad\qquad \delta = \delta_i\nonumber\\
& \qquad\qquad\qquad\qquad \hat{r} = \hat{i}\nonumber\\
& \qquad\qquad K_{\hat{s},n} =  \hat{r}. \nonumber
\end{align}
Thus, every direction $\hat{s}$ in the D2Q[$M\times8$] is described by a sequence of $M$ hoppings $(K_{\hat{s},1},...K_{\hat{s},M})$, where each individual hoping is along one direction $\hat{i}$ selected from the eight basic directions provided by the D2Q8. These are either axial "a" or  diagonal "d" directions.
 
\begin{figure*}
    \centering
    \includegraphics[width=1.0\textwidth]{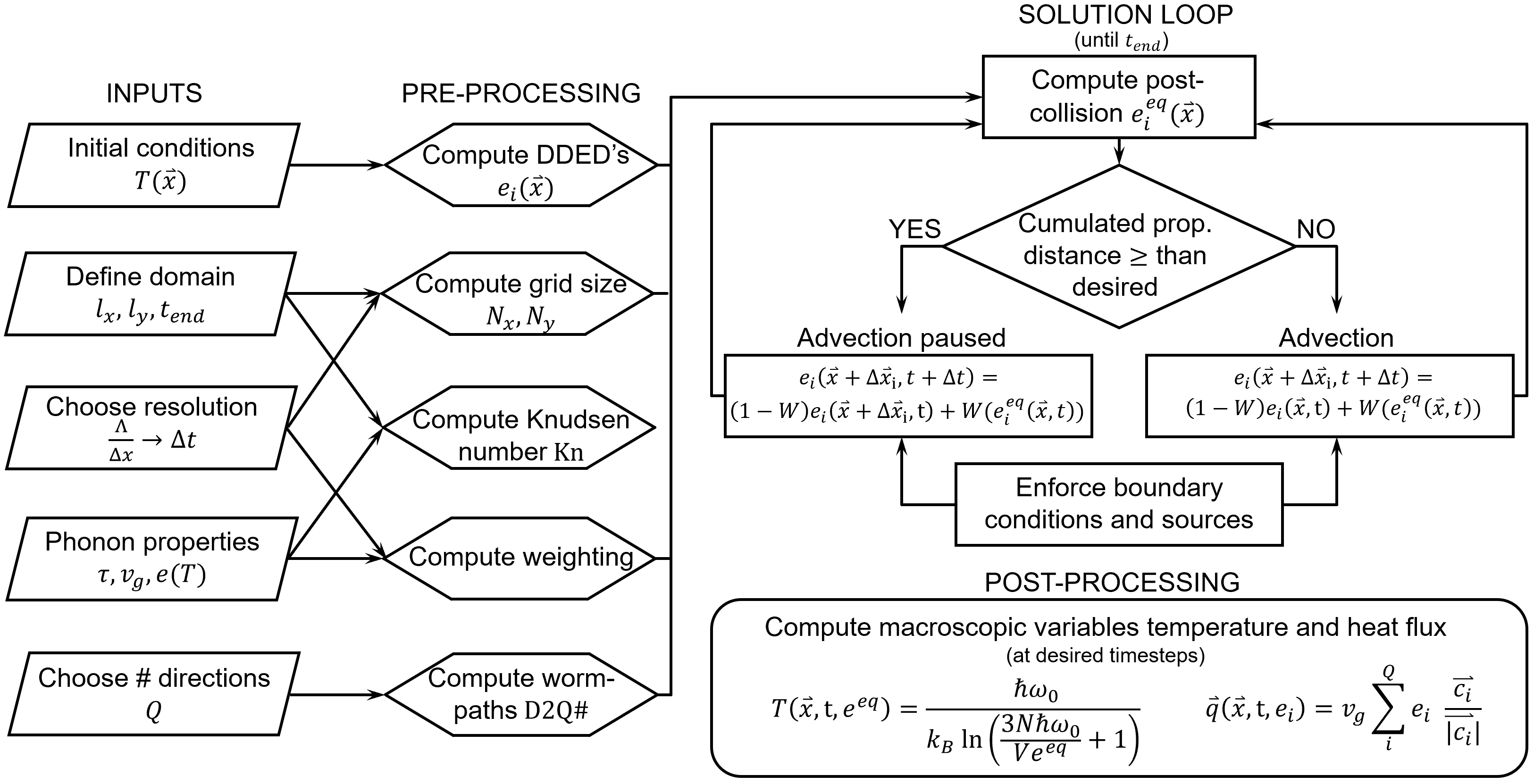}
    \caption{Flowchart of the worm-LBM algorithm}
    \label{fig:wormLBM-flowchart}
\end{figure*}

Figure \ref{fig:CLBM-worm}(b-e) shows the directions (dotted-dashed lines) and the worm trajectories (dashed arrows) for the worm-D2Q16, worm-D2Q24, worm-D2Q32, and worm-D2Q40 schemes. Similar to the TAS-LBM approach, isotropy is enforced by allowing advection only within a radius given by the distance propagated along the axial directions plus one lattice grid spacing. Conversely, scattering  and propagation of the EED contribution is allowed every time step at every lattice site. This method will be, from now on,  referred to as the worm-LBM. Note that the worm-LBM reduces to the TAS-LBM in the limit of $8$ directions. Thus, for the sake of a unified notation, the TAS-D2Q8 will be denoted as the worm-D2Q8 in the following. A flowchart illustrating the worm-LBM algorithm is provided in Fig. \ref{fig:wormLBM-flowchart}.

\subsection{LBM boundary treatment}
\label{sec:boundary}
In the following, the boundary treatment for \textit{isotropic blackbody, anisotropic phonon radiation, adiabatic diffuse, adiabatic specular, and periodic boundary conditions} is provided. It should be remarked that, even if not explicitly stated for the individual cases, the boundary conditions are also allowed to change  their values or character in time.
\\
\\
$\bullet$ \textit{Isotropic blackbody boundary conditions} or, equivalently, isotropic internal blackbody temperature sources at temperature ($T$). These are imposed  by overwriting at every time step the DDEDs ($e_{i}$) by the values defined by the desired temperature. In practice,  at every time step the EED ($e^{eq}$) is distributed according to the angular weighting functions introduced in Eq. \ref{eq:weight}:
\begin{equation}
    e_{i} = w_i e^{eq}(T) \quad \forall, i \in [1,Q],
    \label{eq:isotropic}
\end{equation}
where $e^{eq}(T)$ is computed according to a prescribed temperature value $T$ using Eq. \ref{eq:eq-energy}.
\\
\\
$\bullet$ \textit{Anisotropic phonon radiation boundary conditions} or, equivalently, internal phonon radiation sources. These  are  imposed by overwriting at every time step the DDEDs ($e_{i}$) using a desired radiation distribution $e_{i,\text{imposed}}$:
\begin{equation}
    e_{i} = e_{i,\text{imposed}} \quad \forall \,i \in [1,Q].
    \label{eq:anisotropic}
\end{equation}
\\
\\
$\bullet$ \textit{Adiabatic diffuse boundary conditions} or, equivalently, walls with internal diffuse adiabatic scattering.   These  are generated by updating the DDEDs ($e_{i}$) pointing away from the boundary inwards into the domain,  by the weighted sum of DDEDs ($e_{j}$) pointing towards the boundary:
\begin{equation}
     e_{i} = w_{i} \sum_{j} e_{j} \quad \forall\, i \in \text{domain inwards}.
    \label{eq:adiabatic-diffuse}
\end{equation}
The sum runs over all DDEDs traveling  from the domain towards the boundary.
\\
\\
$\bullet$ \textit{Adiabatic specular boundary conditions} or, equivalently, walls with internal specular adiabatic scattering. These  are generated by substituting the DDEDs ($e_{i}$) pointing away from the boundary inwards into the domain, by their mirror-like equivalents ($e_{j}$) pointing towards the wall, where the mirror plane is the boundary
\begin{eqnarray}
\label{eq:adiabatic-specular}
     e_{i} = e_{j}\quad& \forall\, i \in \text{inwards},&  \forall\, j \in \text{outwards}\\
     &\text{and}& \forall\, (i,j)\ \text{mirrored pairs}.\nonumber
   \end{eqnarray}
\\
$\bullet$ \textit{Adiabatic specular-diffusive boundary conditions}. These can be generated as a linear combination of  Eqs. \ref{eq:adiabatic-diffuse} and \ref{eq:adiabatic-specular}.
\\
\\
$\bullet$ \textit{Flux boundary conditions} or, equivalently, walls with internal flux sources. These are generated by imposing a desired flux vector $\vec{q}$. This boundary condition is not unique, since a specific flux vector can be generated by different combinations of DDEDs. The differences are vanishing in the diffusive limit, but are significant in the ballistic one. Thus, in general, it is recommended to prescribe all the DDEDs uniquely (see "anisotropic phonon radiation boundary conditions"):
\begin{equation}
    \vec{q}= v_\text{g} \sum_{i}^{Q}  e_{i,\text{imposed}} \frac{\vec{c_i}}{|\vec{c_i}|},
    \label{eq:lbm-heat-flux}
\end{equation}
where the sum runs over all $Q$ directions.
\\
$\bullet$ \textit{Periodic boundary conditions}. These are imposed, in 2D, by   linking the leftmost to the rightmost, and the lowermost to the uppermost lattice points.  For instance, a lattice point sitting on the leftmost boundary receives all directions coming from the right of the lattice point periodically located at the rightmost boundary.

\section{Benchmark of the worm-LBM algorithm}
\label{sec:validation}
In the following, the reliability and accuracy of the worm-LBM algorithm is shown. This is done by comparing the ballistic-diffusive phonon transport problem in 2D to reference  solutions (transient and stationary). The latter are either given by analytical solutions, or are obtained from the energy-based variance-reduced Monte Carlo algorithm of Peraud et al. (VRMC) \cite{Peraud-2011} using the program \textsc{Phonon-code} \cite{Phonon-Code}. At this point, it should be emphasized that the worm-LBM algorithm has its strength in the intermediate to high Knudsen number regime. Simulations, aiming for a stationary solution at very low Knudsen numbers (diffusive regime), are numerically very expensive. The diffusive regime can be accurately covered by the Fourier heat equation. Here, the simulations in the very low Kn regime are performed mainly to verify the accuracy of the method by using available analytical solutions. The investigated test cases were:
\begin{itemize}
\item \textit{The transient "one-pixel source"}, which demonstrates the behavior of the method on the grid scale (Sec.  \ref{sec:one-pixel}).
\item \textit{The transient of a uniform circular initial condition}, which demonstrates convergence with the number of propagation  directions. A comparison to the analytical solution in the ballistic regime is also provided  (Sec.  \ref{sec:transient-circle-source}).
\item \textit{The transient of a Gaussian initial condition in the Fourier limit}, which is compared to the simple transient analytical Gaussian solution of the Fourier heat equation (Sec.  \ref{sec:transient-Gaussian}).
\item \textit{The transient cross-plane heat transport}, which is compared to analytical solutions in the ballistic and  Fourier limits, and  numerical solutions obtained in the framework of VRMC (Sec.  \ref{sec:transient-cross-plane}).
\item \textit{The transient in-plane heat transport}, which is compared to the Fuchs-Sondheimer analytical solution for the effective thermal conductivity \cite{Amon2010,Baoling2015} in the steady limit (Sec. \ref{sec:transient-in-plane}).
\item \textit{The stationary one hot - three cold boundaries test-case}, which is compared to the analytical solution in the Fourier limit   (Sec. \ref{sec:stationary-one-hot-three-cold}).
\end{itemize}

All simulations were performed in the gray approximation with the phonon properties reported in Tab. \ref{tab:phonon-properties}.
\begin{table}[h]
    \centering
    \begin{tabular}{|c|c|c|c|c|c}
         $v_g$ & $\tau$ & $\lambda$ & $\omega$ & $\kappa$ & $\alpha$   \\
         (m/s) & (ps)& (nm) & (THz) & (W/(m K) & (cm$^2$/s)  \\
          \hline
          6400 & 6.53 & 41.792 & 81.8 & 129.998 & 0.892\\ 
    \end{tabular}
    \caption{Phonon and  transport properties used for all the simulations reported in this work (MC and worm-LBM). $v_\text{g}$  $\tau$, $\lambda$, $\omega$, $\kappa$, and $\alpha$ denote, respectively, group velocity, phonon lifetime, phonon frequency, thermal conductivity, and diffusivity.}
    \label{tab:phonon-properties}
\end{table}

\subsection{The transient "one-pixel source",}
\label{sec:one-pixel}
The "one-pixel source" consists of one hot lattice point at temperature ($T_\text{hot}$) located at the center of a domain. The latter is initialized homogeneously at a colder temperature ($T_\text{cold}$). The hot lattice point is kept at a constant temperature, using isotropic blackbody boundary conditions. 
 The horizontal and vertical limits of the domain are treated under periodic boundary conditions. A coarse grid (51x51) is chosen  to visualize the  temperature distribution at the grid level. This test is useful to  evaluate the  correctness of the advective paths within the worm-LBM schemes, and is highly recommended to  test the implementation of the worm-LBM method. As this case is the most challenging in terms of ray effect, numerical smearing, and angular false scattering, it also serves to demonstrate that the worm-LBM algorithm overcomes these problems.
 \begin{figure}
    \centering
    \includegraphics[scale=0.5]{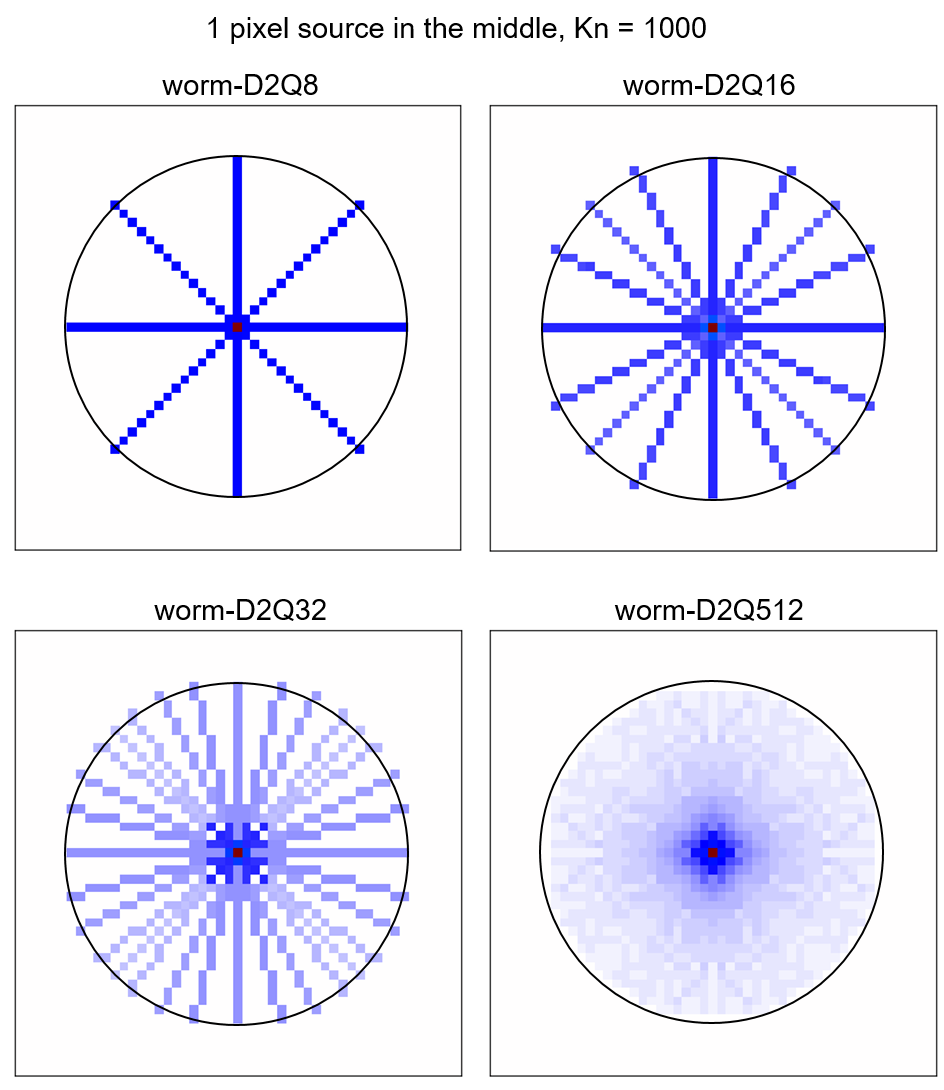}
    \caption{Temperature distribution of "one-pixel source" kept at a constant hot temperature ($T_\text{hot}$) in a domain initialized at a uniform lower temperature  ($T_\text{cold}$) .  The temperature distribution is visualized  after 20 time steps for the worm-D2Q8, worm-D2Q16, worm-D2Q32, and worm-D2Q512.  A coarse grid of 51x51 lattice points is used. }
    \label{fig:one-pixel}
\end{figure}

 As shown in Fig. \ref{fig:one-pixel}, the ray effect is reduced by increasing the number of directions, and disappears for the worm-D2Q512 scheme. The number of directions necessary to overcome the ray effect problem depends on the ratio  between the source size and the perimeter of the circle depicting the "wave front".  Thus, for the extreme case of a one-pixel source,  the scheme becomes approximately angular space filling (ASF) if the number of angular directions fulfills  $Q_{\text{ASF}}\geq n*2\pi$, where $n$ is the number of time steps, which are equivalent to the number of propagated lattice grids along the axial directions.  This can be generalized for a source of diameter $l_\text{source}$, propagating a distance $l_\text{propagation}$:
\begin{equation}
    Q_{\text{ASF}}\geq \frac{l_\text{source}}{l_\text{propagation}}*2\pi.
\end{equation}
It is observed, and it is inherent to the worm-LBM algorithm, that the error in the propagation distance ($x_i$) along a given direction ($i$) is at maximum one pixel for all times:
\begin{equation}
    x_{i,\text{error}}\leq \Delta x_i,
\end{equation}
i.e one grid spacing ($\Delta x_i$) along the propagation direction. The diagonals would be the directions with the maximum possible error. The  axial directions do not show error, since  the pausing mechanism introduced in Eq. \ref{eq:algorithm} does not affect them. Thus, they can be used  as a reference to estimate the error in the distance propagated along the diagonals (see example in Fig. \ref{fig:hopping}). In general, a maximum deviation of one grid spacing perpendicular to the propagation direction is obtained, in agreement with the algorithm encoded for the worm-paths (see Eq. \ref{eq:worm-paths} and Fig. \ref{fig:CLBM-worm}).
 
\begin{figure}[h]
    \centering
    \includegraphics[scale=0.5]{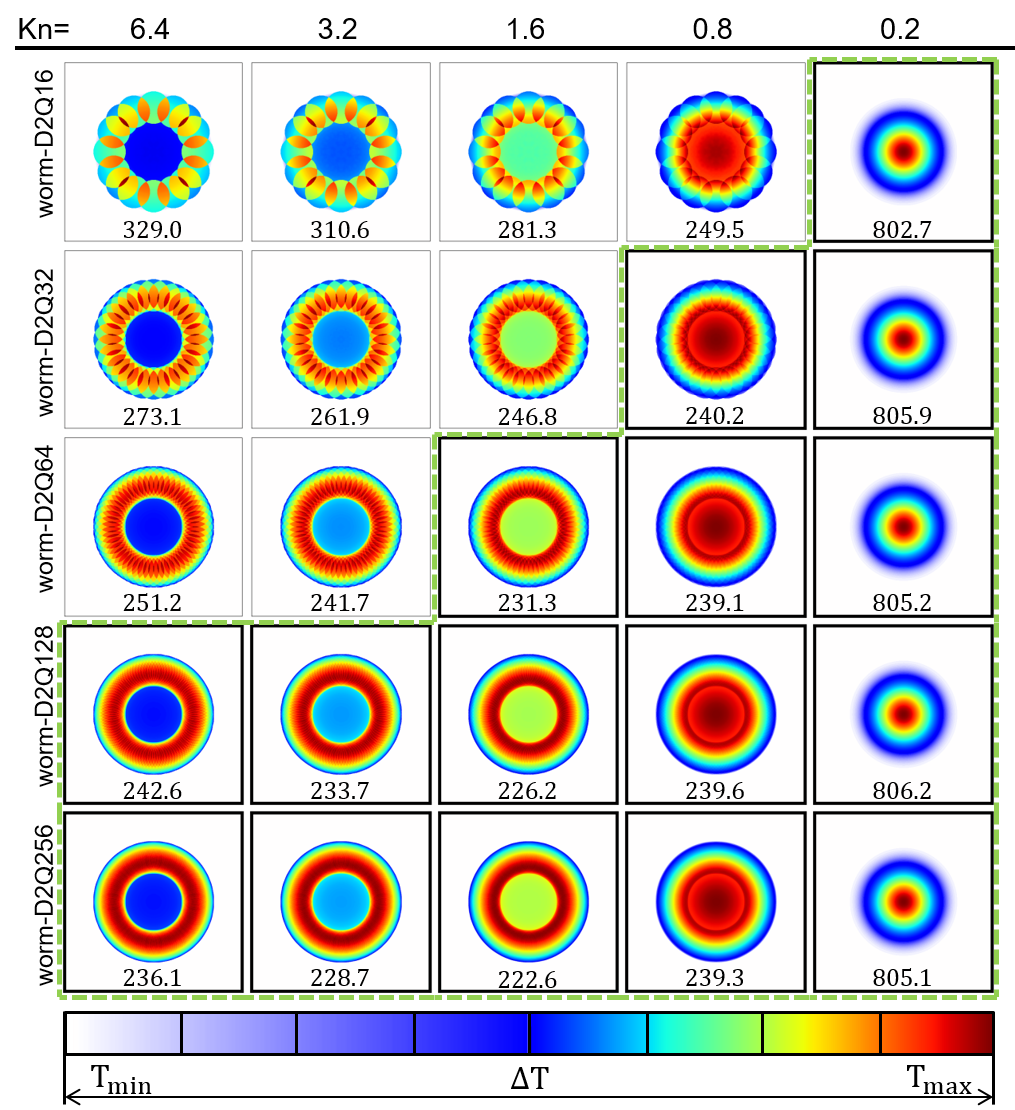}
    \caption{Temperature convergence with the number of directions Q = 16, 32, 64, 128, 256,  for Knudsen numbers Kn = 6.4, 3.2, 1.6, 0.8 and 0.2. A grid of size 512$\times$512 lattice points was used for all simulations. A central circular region of the size 1/11 of the domain size was set  at 301 K, while the domain outside of the circle was initialized at 299 K. The results are shown after 127 time steps. The thick-border black boxes framed with green dashed lines highlight the converged simulations, taking as a criterion a relative error threshold of 3 \% with respect to the solution with the highest number of directions (i.e. the D2Q256). The color map starts a the minimum Temperature $T_{\text{min}} = 299$ K (white), and is re-scaled to the highest temperature occurring in the simulation (dark red). The latter is given by the temperature difference $\Delta T = T_{\text{max}}-T_{\text{min}}$ in mK as shown below each sub-figure.}
    \label{fig:Q-convergence}
\end{figure}
\subsection{The transient of a uniform circular initial condition}
\label{sec:transient-circle-source}
Although the transient uniform circular source condition is related to the "one-pixel" problem studied before, it is more practical and closer to a real physical situation.  Here, a hot circular initial condition with uniform temperature $T_\text{hot}=301$ K and radius 1/11 of the domain size is initialized in a uniform cold background at $T_\text{cold}=299 \,\text{K}$.   Periodic boundary conditions are imposed on the horizontal and vertical edges of the domain (see Sec. \ref{sec:boundary}). This test case is used to quantify the convergence with the number of propagation directions for Knudsen numbers in the ballistic and diffusive regime. As shown in Fig. \ref{fig:Q-convergence}, due to the ray effect, the solution does not converge for a too low number of lattice directions.  The thick-border black boxes, framed by green dashed lines, highlight the simulations that are within 3 \% of the error for the highest temperature in the domain. The error is calculated, taking as reference the solution obtained with the worm-D2Q256 scheme, which has the highest number of propagation directions among the schemes considered. It can be seen that higher Knudsen numbers require  much more directions to reach convergence.
\begin{figure}[h]
    \centering
    \includegraphics[width=0.5\textwidth]{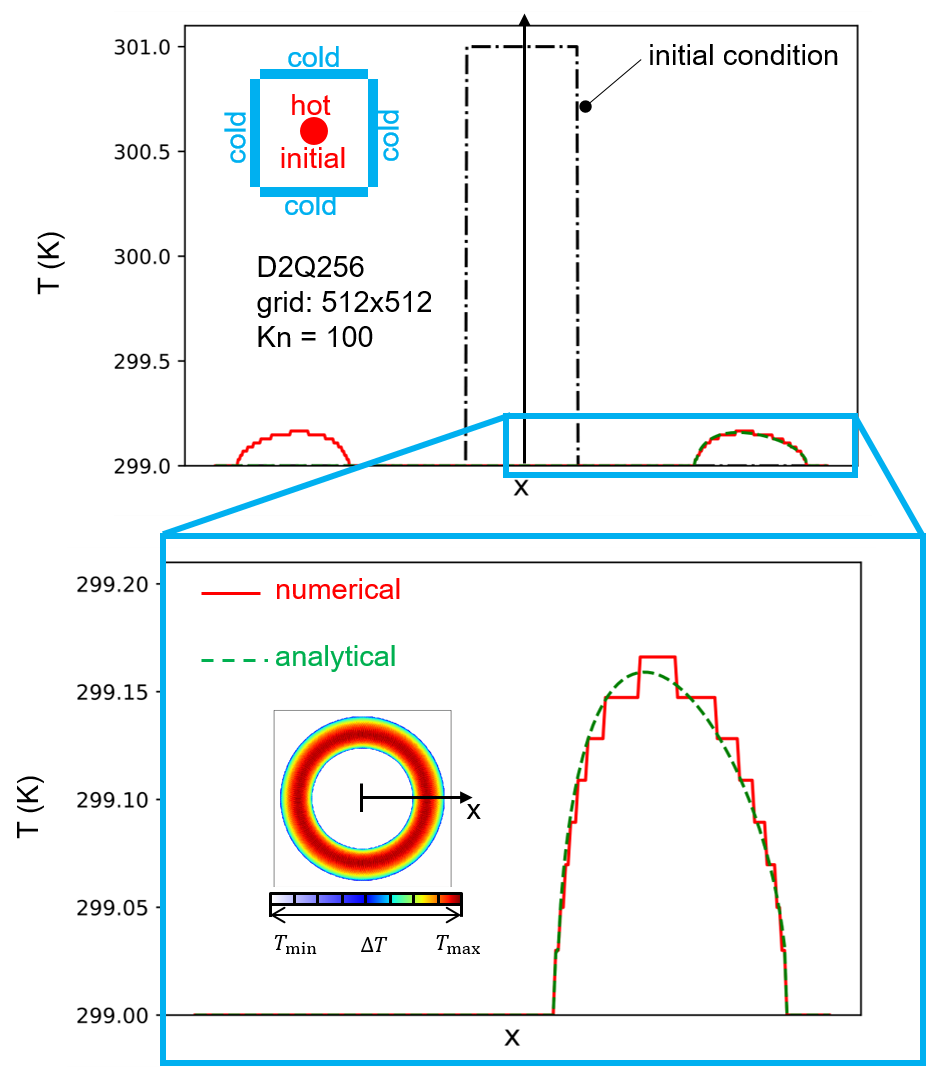}
    \caption{Temperature profile of a uniform circular initial condition:  comparison of the worm-D2Q265 algorithm to the analytical solution in the the ballistic limit   ($\text{Kn} = 100$).  A hot circular initial condition at $T_\text{hot}=301$ K and radius 1/11 of domain size, is initialized in a uniform cold background at $T_\text{cold}=299$ K (dot-dashed black line). The solution is shown after 192 time steps (red and green curves), which corresponds to $t\approx 65.3$ fs. A grid with 512x512 lattice points was used. The temperature 
 profile was calculated along the $x$ direction for $y=Ny/2$. The inset in the lower panel shows the temperature distribution at $t\approx 65.3$ fs.}
    \label{fig:circle-ballistic-analytic}
\end{figure}

In this particular case, it is straight forward to derive an analytical solution for the resulting intensity profile in the ballistic limit. The intensity (or temperature) along the x-axis can be computed by the circle equation $(x-v_\text{g}t)^2+y^2=R^2$, weighting it by the circle density $1/x$, where $x$ is the distance propagated from the midpoint. Since the problem fulfills rotational symmetry, the solution can be written for an arbitrary direction as:
\begin{equation}
    T(r,t) = \frac{T}{\pi}\frac{\sqrt{R^2-(r-v_\text{g}t)^2}}{r}, 
    \label{eq:circle-analytic}
\end{equation}
where $r=|\vec{r}|$ is the distance from the midpoint. 

Figure \ref{fig:circle-ballistic-analytic} shows a comparison to the analytical solution for the temperature distribution obtained with the worm-D2Q256 scheme, and $\text{Kn} = 100$.   The solution is reported after propagating 192 time steps (i.e. $t\approx 65.3$ fs). The influence of the angular discretization is still slightly visible for this very high Knudsen number case.  A stepped appearance in the temperature profile can be seen by zooming into the profile (lower-panel  Fig. \ref{fig:circle-ballistic-analytic}). 

\subsection{The transient of a Gaussian initial condition in the Fourier limit}
\label{sec:transient-Gaussian}
\begin{figure}[h]
    \centering
    \includegraphics[scale=0.55]{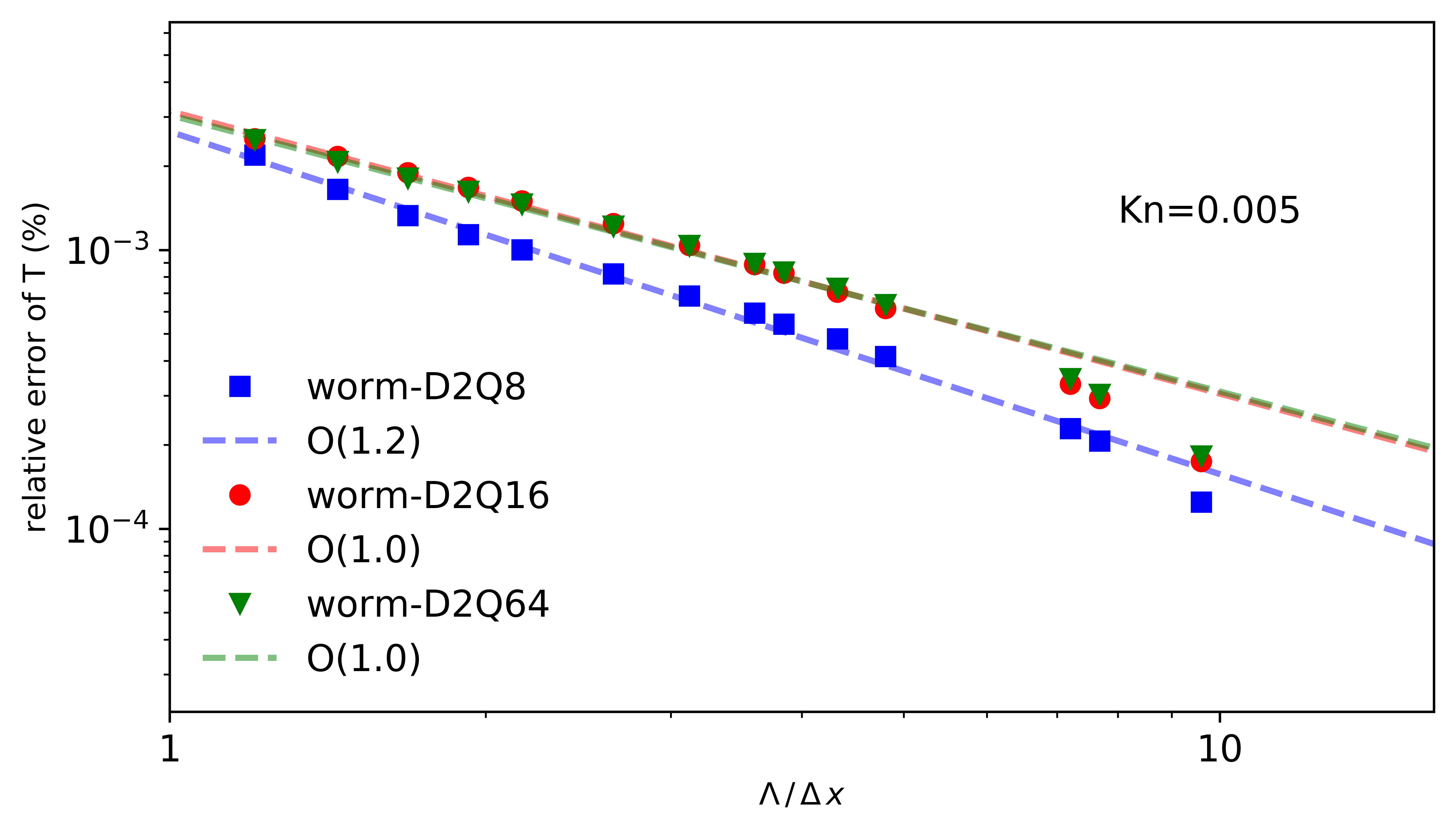}
    \caption{Grid convergence of the worm-D2Q8, worm-D2Q16 and worm-D2Q64 schemes for a  Gaussian initial condition, and  Kn$=0.005$. The temperature value in the middle of the domain (hottest point) is compared to the value obtained with the finest grid calculated ($\lambda/\Delta x=14.4$).  The data  correspond to  time $t=195.9 \,\tau\approx1279$ ps. The dashed lines represent different convergence tendencies obtained from fitting the numerical data.   Table \ref{tab:gauss} list the parameters  used for the Gaussian initial distribution (Eq. \ref{eq:Gaussian}).}
    \label{fig:grid-convergence-Gaussian}
\end{figure}

\begin{figure}[h!]
\centering
\includegraphics[scale=0.45]{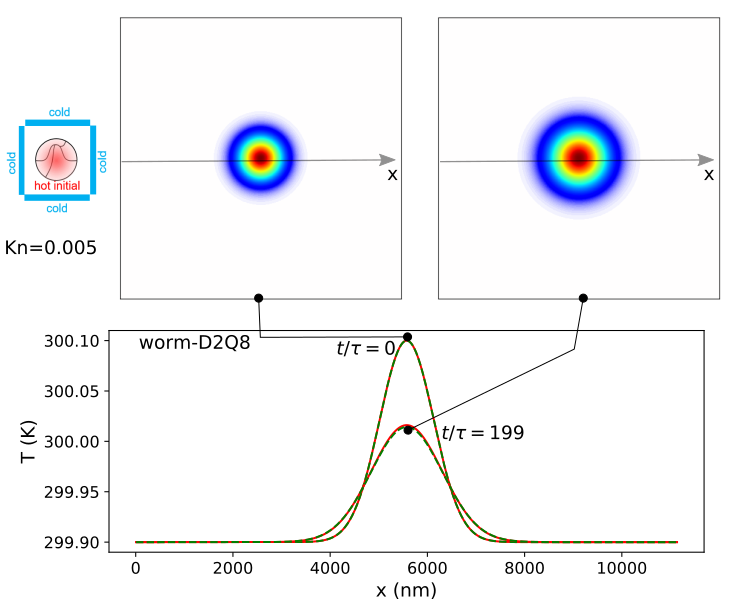}
    \caption{Comparison of the transient temperature profiles obtained with the worm-D2Q8 (red solid line) to the analytical solution (green dashed line) for a Gaussian initial condition (black dash dotted line). The temperature profile corresponds to  $t=199.8 \tau\approx1305$ ps, and $\text{Kn}=0.005$. Moderately fine grid resolution  was used, such that  $1/W=\tau/\Delta t = \text{MFP}/\Delta x = 6$ ($N_x \times N_y =1600\times1600$). Table \ref{tab:gauss} list the parameters  used for the Gaussian initial distribution (Eq. \ref{eq:Gaussian}). The temperature 
    profiles were calculated along the $x$ direction for $y=N_y/2$. The upper panels show the temperature distributions in the entire domain at different time steps.}
    \label{fig:gaussian}
\end{figure}

In this test case, the worm-LBM is compared to the 2D analytical solution for the temperature distribution of a  Gaussian initial condition in the Fourier limit. The analytical solution for the temperature, as derived from the heat equation,  is \cite{wiki:Heat_equation}:
\begin{equation}
    T(x,y,t) = \frac{4\sqrt{\alpha\,t_0}}{\sqrt{\alpha\,(t_0+t)}} e^{-\frac{(x-x_0)^2+(y-y_0)^2}{4\,\sqrt{\alpha\,(t_0+t)}}},
    \label{eq:Gaussian}
\end{equation}
with $t_0=\sigma_0^2/(2k)$. Here, $\sigma_0$ is the initial width defined in terms of the Gaussian standard deviation, and $\alpha$ is the phonon thermal diffusivity. 
Figure \ref{fig:grid-convergence-Gaussian} 
shows the grid convergence of  the worm-D2Q8, worm-D2Q16 and worm-D2Q64 schemes for a  Gaussian initial condition, and  $\text{Kn}=0.005$. A first-order convergence rate is observed, and  the coarser grid in all cases shows errors below 0.004 \% with respect to the solution obtained with the finest grid considered. As a high number of directions is not necessary in the diffusive limit, the worm-D2Q8 is already in excellent agreement with the analytical solution (Fig. \ref{fig:gaussian}).
\begin{table}[h!]
    \centering
    \begin{tabular}{|c|c|c|c|c}
         $x_0$ (nm)& $y_0$ (nm)&$\sigma_0$ &$k$  ($\text{cm}^2$/s)  \\
          \hline
          5567.63& 5567.63&556.76&0.89\\ 
    \end{tabular}
    \caption{Parameters for the Gaussian initial distribution (Eq. \ref{eq:Gaussian}) used in our simulations. }
    \label{tab:gauss}
\end{table}

\subsection{The transient cross-plane heat transport}
\label{sec:transient-cross-plane}
\begin{figure}[h]
    \centering
    \includegraphics[scale=0.5]{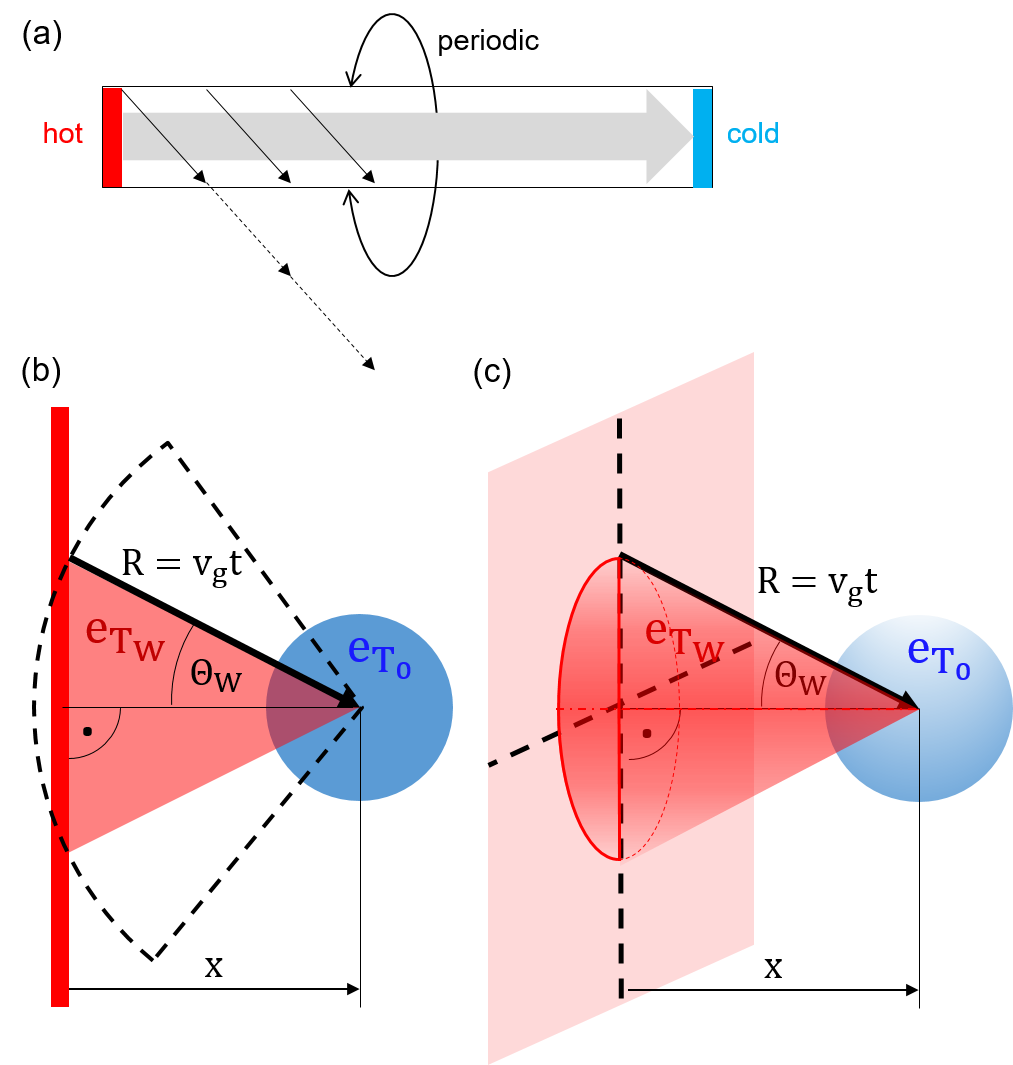}
    \caption{(a) Schematic of the cross-plane test case.  (a) Planar angle in 2D, and (c) solid angle in 3D measured from a point located at a distance $x$ to the wall.}
    \label{fig:transient-1D-2D-3D}
\end{figure}

The transient cross-plane heat transport corresponds to a 1D problem with periodic boundary conditions perpendicular to the heat flux (see Fig. \ref{fig:transient-1D-2D-3D}a).  Isotropic blackbody boundary conditions with temperatures $T_h=300.5$ K and  $T_c=300$ K are imposed on the left and right walls, respectively. The domain is initialized at $T_c=300$ K. 

In the diffusive limit ($\text{Kn}<<1$), the temperature and heat flux distributions are compared to the analytical solutions \cite{GUO20161}:
\begin{eqnarray}
    \Theta(X,\xi) &=&\\\nonumber
    1 \,-\, X &-& \frac{2}{\pi} \sum_{l=1}^\infty \frac{1}{l}\sin{(l\pi X)}\exp{\left(-\text{Kn}\frac{l^2\pi^2\xi}{3}\right)}
    \label{eq:1DT}
\end{eqnarray}
and
\begin{equation}
    Q_x(X,\xi) = 1 + 2\sum_{l=1}^\infty \frac{1}{l}\cos{(l\pi X)}\exp{\left(-\text{Kn}\frac{l^2\pi^2\xi}{3}\right),}
    \label{eq:1Dhx}
\end{equation}
with
\begin{eqnarray}
   & X = \frac{x}{L},\;\; \xi = \frac{t}{\tau}\,\text{Kn},  \;\Theta = \frac{T-T_c}{T_h-T_c} &\\ \nonumber &\text{and} \; Q_x=\frac{q\,L}{\kappa (T_c-T_h)}.&
    \label{eq:1D-norm-varibles}
\end{eqnarray}

\begin{figure}[h]
    \centering
    \includegraphics[scale=0.5]{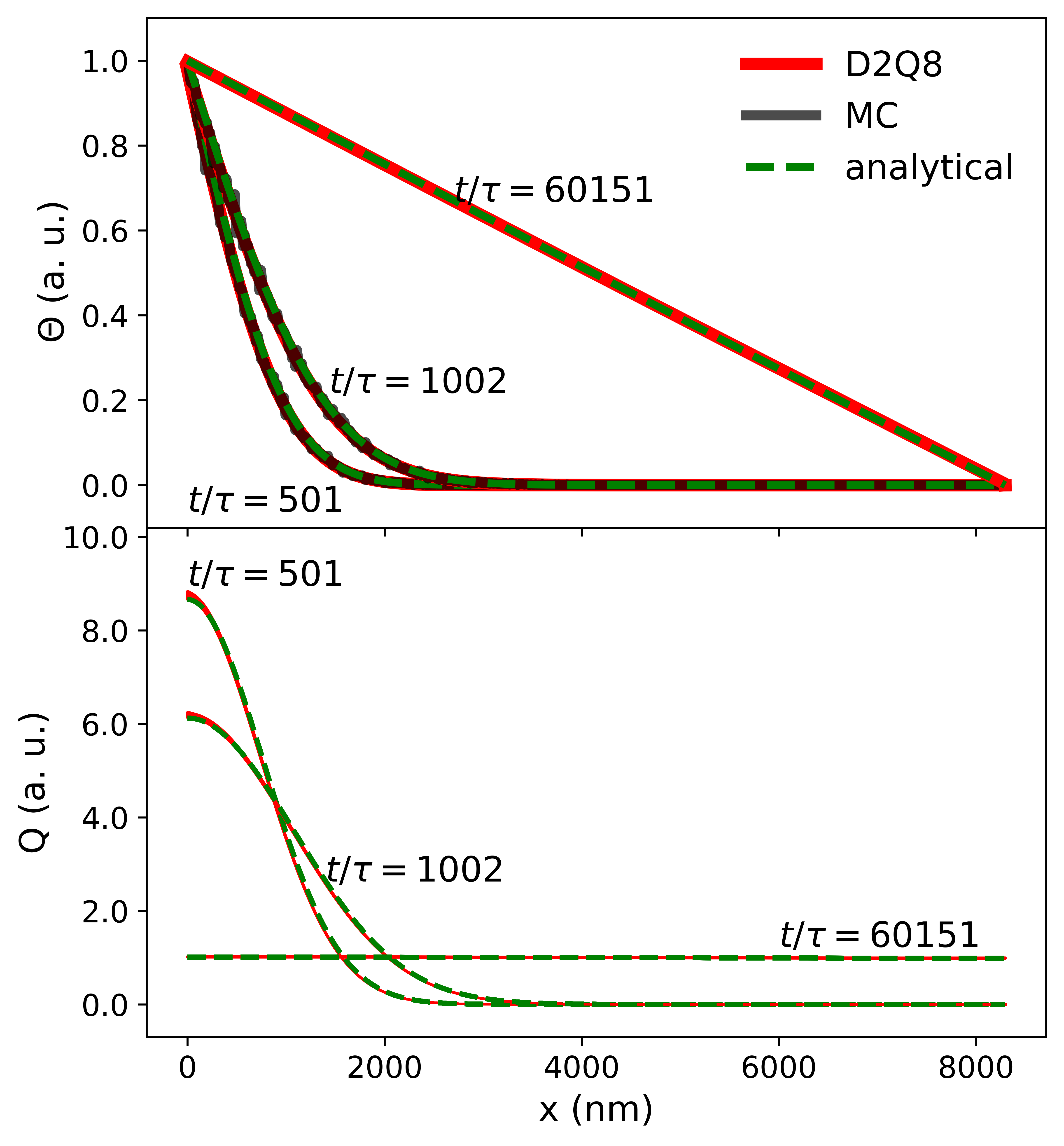}
    \caption{Fourier cross-plane temperature and heat profiles predicted by the worm-D2Q8 algorithm compared to the analytical solutions (Eqs. \ref{eq:1Dhx} and \ref{eq:1DT}).  The MC results were calculated using the \textsc{phonon-code} \cite{Phonon-Code}. $T_h= 300.5$ K, $T_c= 300.0$ K, $N_x=12670$, $N_y=1$,   Kn=0.005, $1/W\approx63.8$. At every time step, $t/\tau$, reported in the lower-panel, heat fluxes for $t/\tau \pm 10$ are shown. The transient MC simulations were performed using 5000000 particles, 50000 maximum number of scattering events, and a grid of 500$\times$1 detectors (see details about the parameters in \cite{Phonon-Code}).The steady MC was no calculated for this Knudsen number.  }
    \label{fig:cross-plane-Fourier}
\end{figure}

Here, $L$ is the non-normalized domain length in the heat flux direction, and $q$ is the magnitude of the non-normalized heat flux.
As it can be seen in Fig. \ref{fig:cross-plane-Fourier}, the transient and the steady states of the cross-plane heat transport problem in the Fourier limit are perfectly described by the new worm-LBM algorithm. There is a nearly perfect agreement between the  worm-LBM results, the analytical solutions and the MC results.

\begin{figure}[h!]
    \centering
    \includegraphics[scale=0.5]{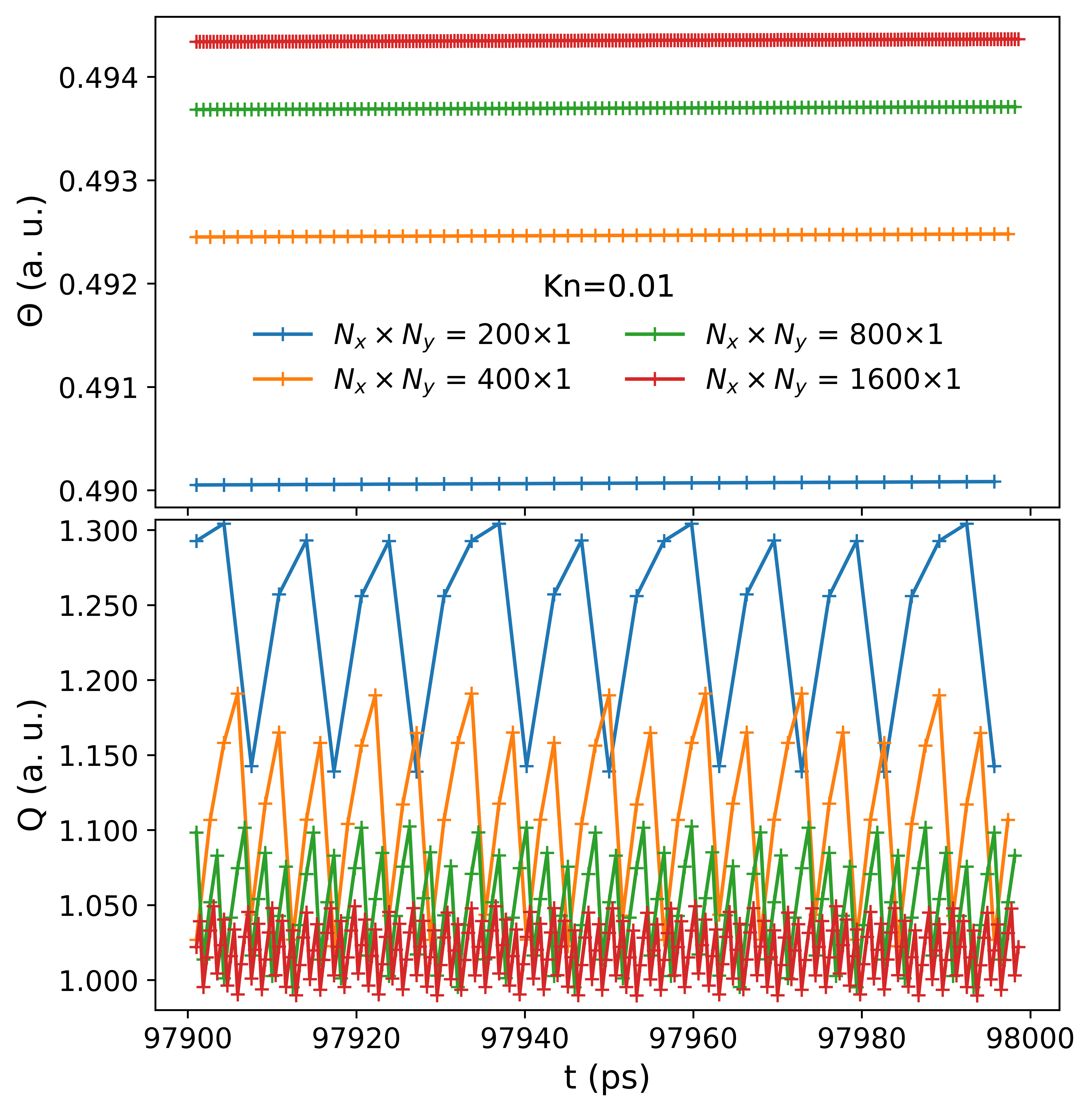}
    \caption{Time evolution of the effective temperature and heat flux of the cross-plane heat transport at the center of the domain ($N_x/2$)  close to the steady state. As values are normalized according to Eq. \ref{eq:1D-norm-varibles}, and are reported close to the steady state, the heat flux and temperature converges with the grid size toward 1 and 0.5, respectively.}
    \label{fig:oscillations}
\end{figure}

The worm-algorithm by pausing the advection of the DDEDs along the diagonal directions, while continuing propagation the EED, introduces small fluctuations in time for the energy. The impact of such fluctuations is negligible for the effective temperature (upper-panel Fig. \ref{fig:oscillations}). Still, they can result in variations of the heat flux of the order of 10 \% for a very coarse grid (lower-panel Fig. \ref{fig:oscillations}). Nonetheless, the amplitude of such fluctuations,  and hence the error of the heat flux, reduces with the grid size (lower-panel Fig. \ref{fig:oscillations}). In practice, a good estimator of the heat flux is an average value (e.g. over 10 time steps) around the desired time. As it can be seen in the example of   Fig. \ref{fig:cross-plane-Fourier},  where the heat flux for ten time steps around the desired times are reported, the fluctuation of the heat flux is small for a well grid converged system. In the following, all heat flux data reported or used to calculate another quantity (e.g. thermal conductivity) is averaged over ten time steps around the time of interest. 

In the ballistic limit ($\text{Kn}>>1$),  the temperature distribution can be computed analytically by combining the discrete energy density coming from the wall, $e_{T_\text{W}}$, and the isotropic contribution defined by the initial condition $e_{T_0}$. $T_\text{W}$ and $T_0$ are, respectively, the wall and initial temperature conditions.  At a given time, only discrete energy densities from a distance $R=v_g \,t$ can reach a point located at a distance $x$ to the wall (Fig. \ref{fig:transient-1D-2D-3D}b). This point will be reached only by energy densities emitted by the wall  within the angular interval $(-\theta_\text{W},\theta_\text{W})$, with $\theta_\text{W}=\arccos{x/R}$. Thus, the total equilibrium energy density can be written as:
\begin{eqnarray}
    e^{eq} &=& 2 \left[ \int_0^{\theta_\text{W}} e_{T_\text{W}} d\Theta + \int_{\theta_\text{W}}^\pi e_{T_0} d\Theta \right]\nonumber\\
    &=& 2\pi e_{T_0} + 2\Theta_\text{W} \left[e_{T_\text{W}}-e_{T_0}\right],
    \label{eq:1D-ball-2D}
\end{eqnarray}
and the temperature can be computed using  Eq. \ref{eq:pseudo-T}.
\begin{figure}
    \centering
    \includegraphics[scale=0.5]{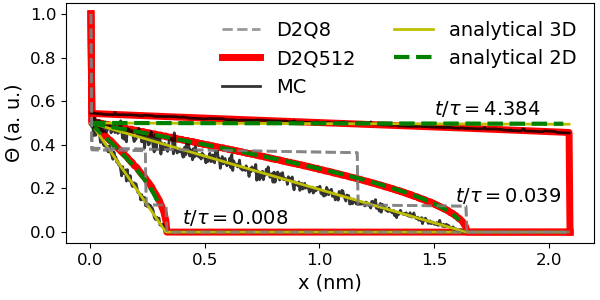}
    \caption{Ballistic cross-plane temperature profile predicted by the worm-D2Q512 algorithm compared to the analytical solution (Eq. \ref{eq:1D-ball-2D} ). $N_x=510$, $N_y=1$,  \text{Kn}=20, and $1/W\approx10225$. The transient and steady MC simulations were performed using 9000000 particles, 90000 maximum number of scattering events, and a grid of 500$\times$1 detectors (see details about the parameters in \cite{Phonon-Code}).  }
    \label{fig:cross-plane-ballistic}
\end{figure}
Fig. \ref{fig:cross-plane-ballistic} shows a comparison of the worm-D2Q512 solutions to the 2D analytical ones (Eq. \ref{eq:1D-ball-2D}), and to results obtained with MC. The MC result gives a linear "wave front" profile as opposed to the other two solutions (2D analytical and worm-D2Q512), which show a temperature profile which is dominated by the $\arccos$ function. Nonetheless, in the limit of long simulation times (once the wave front reaches the left wall) all solutions converge toward the same steady state (e.g. Fig. \ref{fig:cross-plane-ballistic} $t/\tau=4.384$). The difference in the transient regime  has its origin in the dimensionality of the problem. Although the code used in our MC simulations solves the BTE in 2D \cite{Phonon-Code},  the total equilibrium energy density is calculated considering the solid angle (Fig. \ref{fig:transient-1D-2D-3D}c) to emulate the 3D behavior \cite{Peraud-2011}. Thus, the MC temperature profile follows the corresponding  3D analytical solution (Fig. \ref{fig:cross-plane-ballistic}).

For the sake of comparison, Fig.  \ref{fig:cross-plane-ballistic} includes the temperature profile provided by the worm-D2Q8. As can be seen, such a reduced number of propagation directions in the ballistic regime results in a very bad resolution for the temperature distribution.  

\begin{figure}[h]
    \centering
    \includegraphics[scale=0.52]{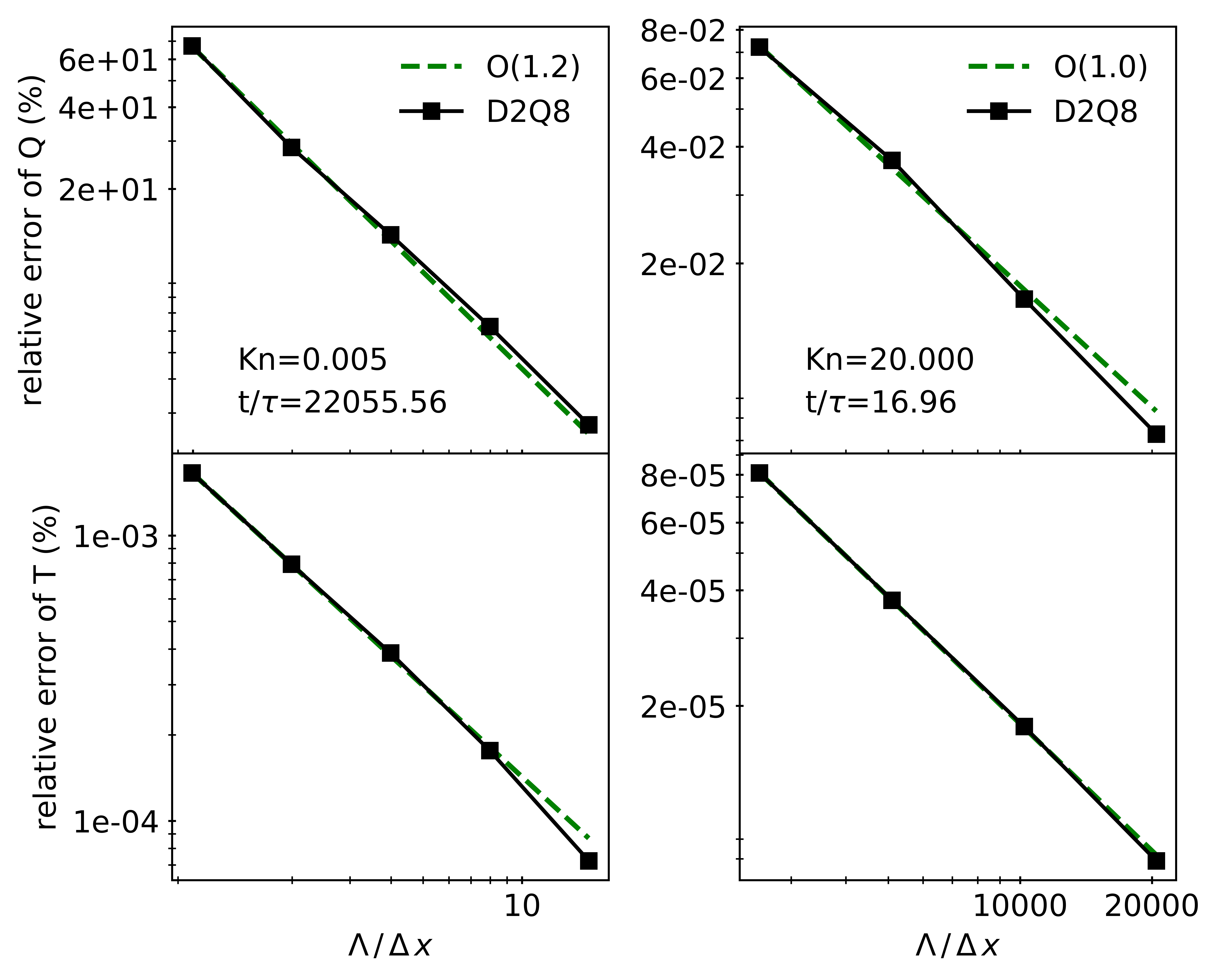}
    \caption{Grid convergence for the transient cross-plane temperature profile. Left panel: $Kn=0.005$. Right panel: $Kn=20$. 
    The temperature and heat flux over the entire domain at a given time-step is averaged and compared to to the numerical solution of the finest grid. The finest used grids were $\lambda/\Delta x\sim64$ ($Kn=0.005$), and $\lambda/\Delta x=81900$ ($Kn=20$).}
    \label{fig:grid-convergence-1D-Q}
\end{figure}

In this test-case, as in the Gaussian initial condition, a first-order convergence rate is observed (Fig. \ref{fig:grid-convergence-1D-Q}). This demonstrates that the order of convergence is preserved in the presence of the applied boundary conditions.   

\begin{figure}[h]
    \centering
    \includegraphics[scale=0.55]{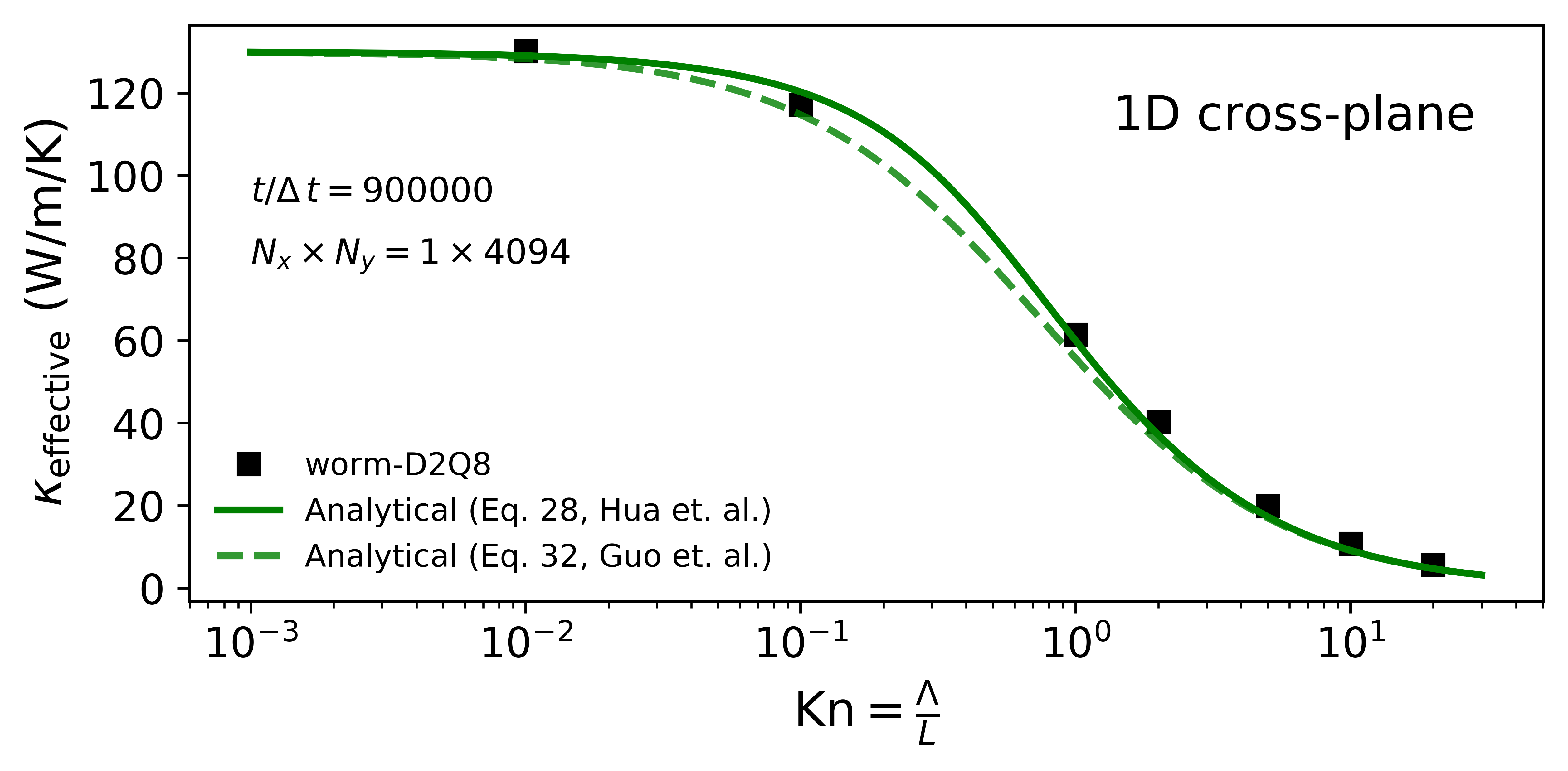}
    \caption{Knudsen number dependence of the 1D cross-plane thermal conductivity at 300.25 K.   Analytical solutions were taken from Hua et.al. \cite{Hua2015}, and Guo et.al. \cite{GUO20161}.}
    \label{fig:kappa-1Dcross}
 \end{figure}

From a practical point of view, it is relevant  to compute the effective 1D cross-plane thermal conductivity for different Knudsen numbers. Figure \ref{fig:kappa-1Dcross} shows a comparison of the worm-D2Q8 thermal conductivities to the analytical solutions provided by Hua et.al. ~\cite{Hua2015} and Guo et.al. ~\cite{GUO20161}. In the entire range of Knudsen numbers, the worm-LBM provides solutions in agreement with the analytical solutions. For low Knudsen numbers, the effective thermal conductivity converges to the bulk thermal conductivity. In practice, the worm-LBM should be used at intermediate and high Knudsen numbers, whereas it is sufficient and computationally more efficient to solve the Fourier heat equation for low Knudsen number dominated problems.   
\subsection{The transient in-plane heat transport}
\label{sec:transient-in-plane}
The transient in-plane and cross-plane heat transport problems are very similar to each other.  The difference relies on the type of  boundary condition applied perpendicular to the main heat flux direction, which in this case is adiabatic-diffuse (see Sec. \ref{sec:boundary}). Moreover, the Knudsen number is defined by the width of the domain, instead of the distance between the hot and cold walls. All the other parameters are kept equal, i.e. isotropic blackbody boundary conditions for the left ($T_h=300.5$)  and right ($T_c=300$) walls, and uniform initial domain temperature ($T_c=300$ K).  After reaching steady state, the numerical solutions for the in-plane thermal conductivity are compared to the Fuchs-Sondheimer analytical expression \cite{GUO20161}. 
 \begin{figure}[h]
    \centering
    \includegraphics[scale=0.5]{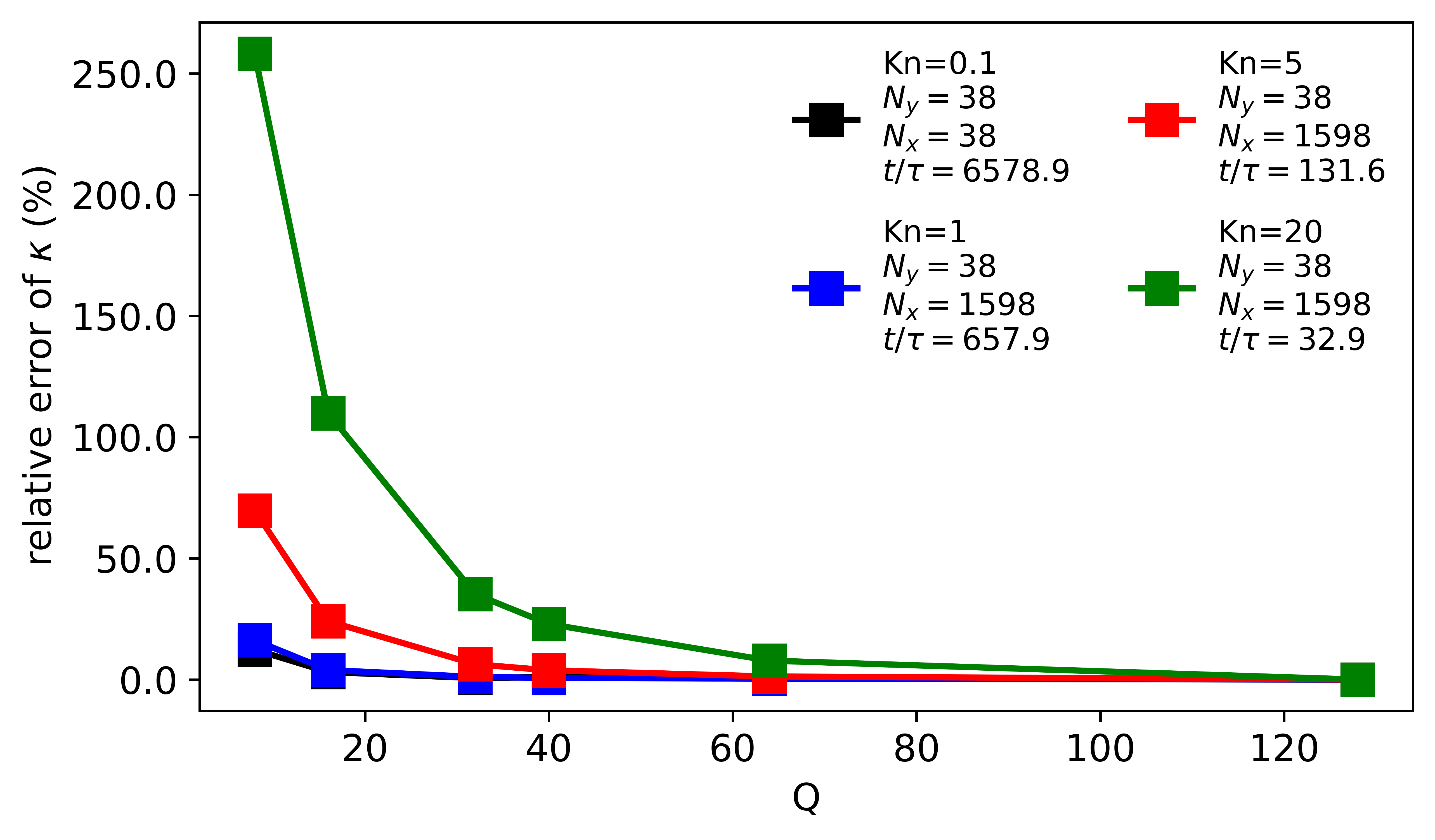}
    \caption{Number of directions convergence (Q-convergence) of the in-plane thermal conductivity, taking as reference the value obtained with the worm-D2Q128 scheme.}
    \label{fig:Q-conv-inplane}
\end{figure}
\begin{figure}[h]
    \centering
    \includegraphics[scale=0.5]{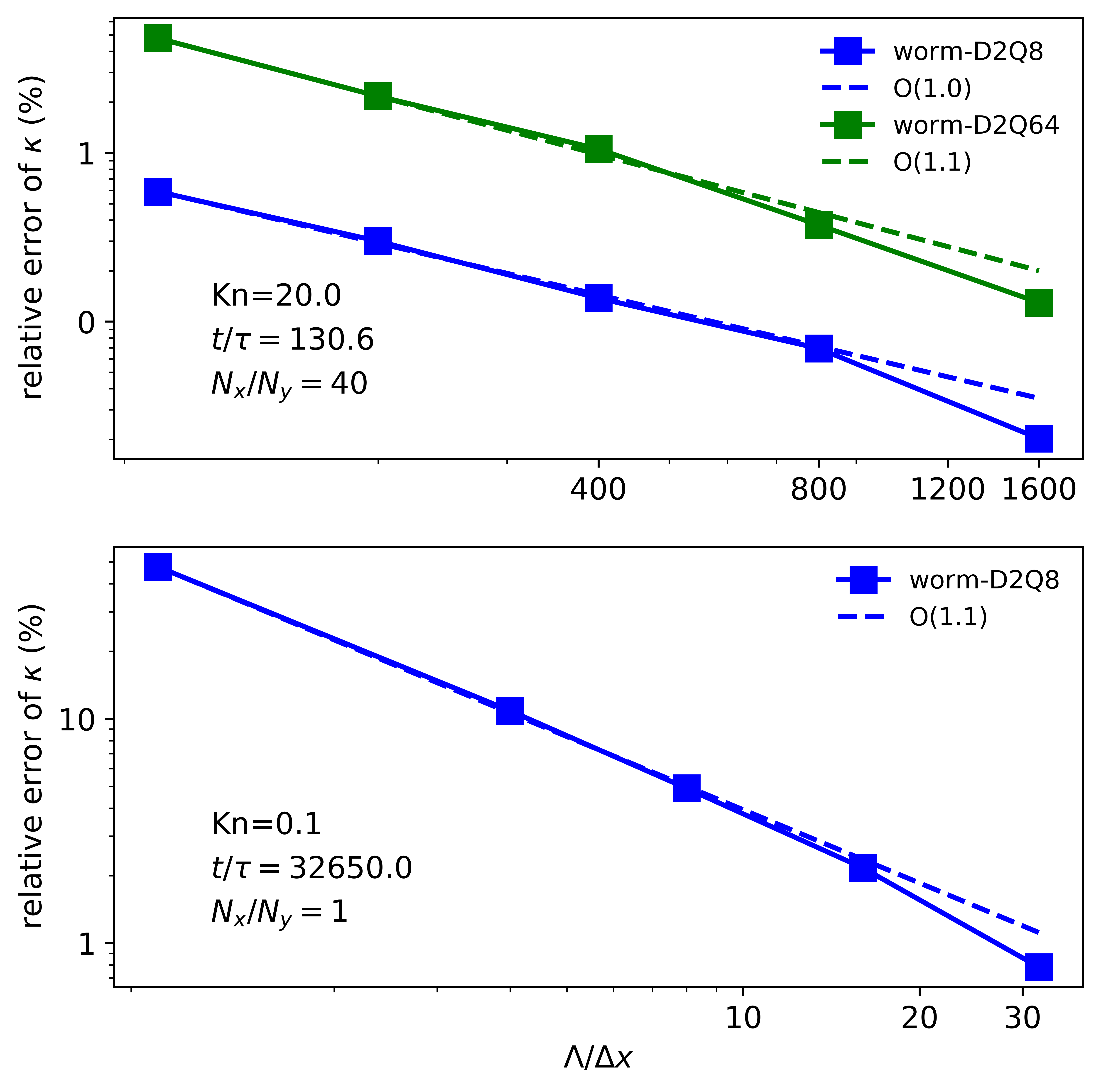}
    \caption{Grid convergence of the in-plane transient thermal conductivity, using as reference the value calculated with the finest grid (i.e. $\lambda/\Delta\,x=3200$ for $\text{Kn}=20$, and $\lambda/\Delta\,x=64$ for $\text{Kn}=0.1$, respectively).}
    \label{fig:Nx-conv-inplane}
\end{figure}
Figure \ref{fig:Q-conv-inplane} shows the convergence in terms of the number of directions (Q-convergence) of the in-plane thermal conductivity in the ballistic and diffusive regimes, taking as a reference the values predicted by the worm-D2Q128 scheme. In the limit of very low Knudsen numbers (diffusive limit) the worm-D2Q8 scheme is  reasonably well converged, with errors of $\sim$ 12\% with respect to the worm-D2Q128 value. Conversely, for high Knudsen numbers the worm-D2Q8 value is much higher than the result obtained with the worm-D2Q128 ($\sim$ 260 \%).  In the D2Q8 scheme, there are just three directions (two diagonals and one axial) available to propagate the energy along the heat flux direction, and a large proportion of the energy is propagated along the axial direction, without being affected by the boundary scattering.   As the number of directions increases,  more and more energy is propagated along directions for which boundary scattering becomes possible. 

Figure \ref{fig:Nx-conv-inplane} shows the grid convergence tendency for the in-plane transient effective thermal conductivity, taking as a reference the value obtained with the finest grid. Here again, a first-order convergence rate is observed.
\begin{figure}[h]
    \centering
    \includegraphics[scale=0.55]{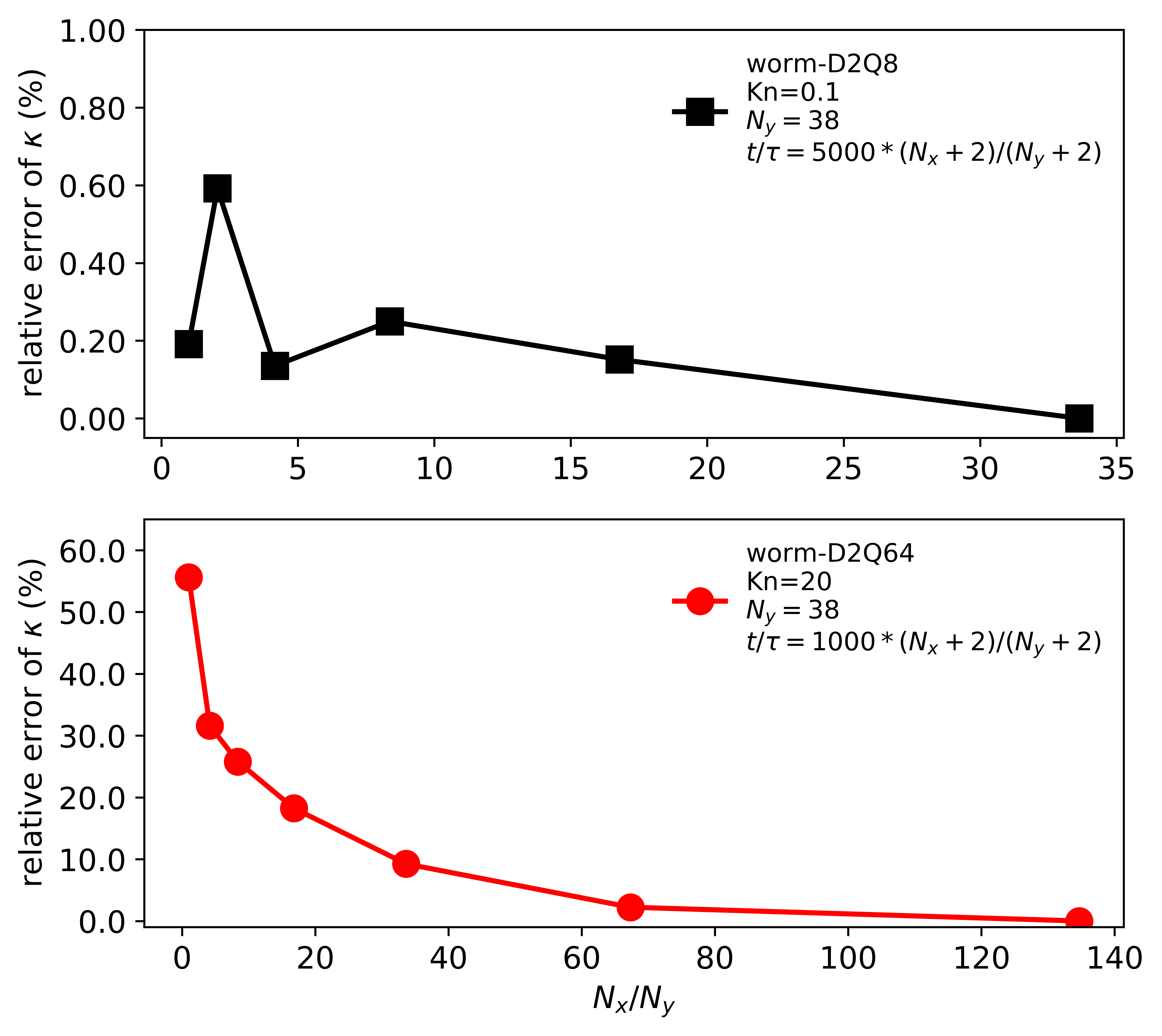}
    \caption{Domain length convergence of the in-plane transient thermal conductivity. $k_\text{ref}$ is the value calculated with the largest domain. The domain length corresponds to the distance between the cold and hot walls (i.e. $N_x\times\Delta x$),  where  the grid space $\Delta\, x$ is defined from the domain width $N_y=38$. }
    \label{fig:ballistic-conv-inplane}
\end{figure}

An important parameter to correctly estimate the in-plane thermal transport in the ballistic regime is the domain length-width ratio. This parameter is independent of the numerical algorithm for solving the BTE, and  has to be considered in any case. In the limit of high Knudsen numbers, one should ensure large ratios to avoid constraints of the phonons mean free path by the hot and the cold walls. Here,  for the highest Knudsen number considered, a ratio of  $N_x/N_y=16.79$ underestimates the value obtained with $N_x/N_y=134.68$ by $\sim$ 18 \% (lower-panel Fig. \ref{fig:ballistic-conv-inplane}). Conversely, in the diffusive limit differences below 1 \% are obtained. The small differences among different $N_x/N_y$ ratios in this case  have origin on the time step at which the solution is computed and compared (see Fig. \ref{fig:ballistic-conv-inplane}), which have not been large enough to fully reach the steady state. Simulation times were chosen following a linear dependence with the length-width ratio, $t/\tau=5000*(N_x+2)/(N_y+2)$, which is not fulfilled in the diffusive regime. 
\begin{figure}[h]
    \centering
    \includegraphics[scale=0.55]{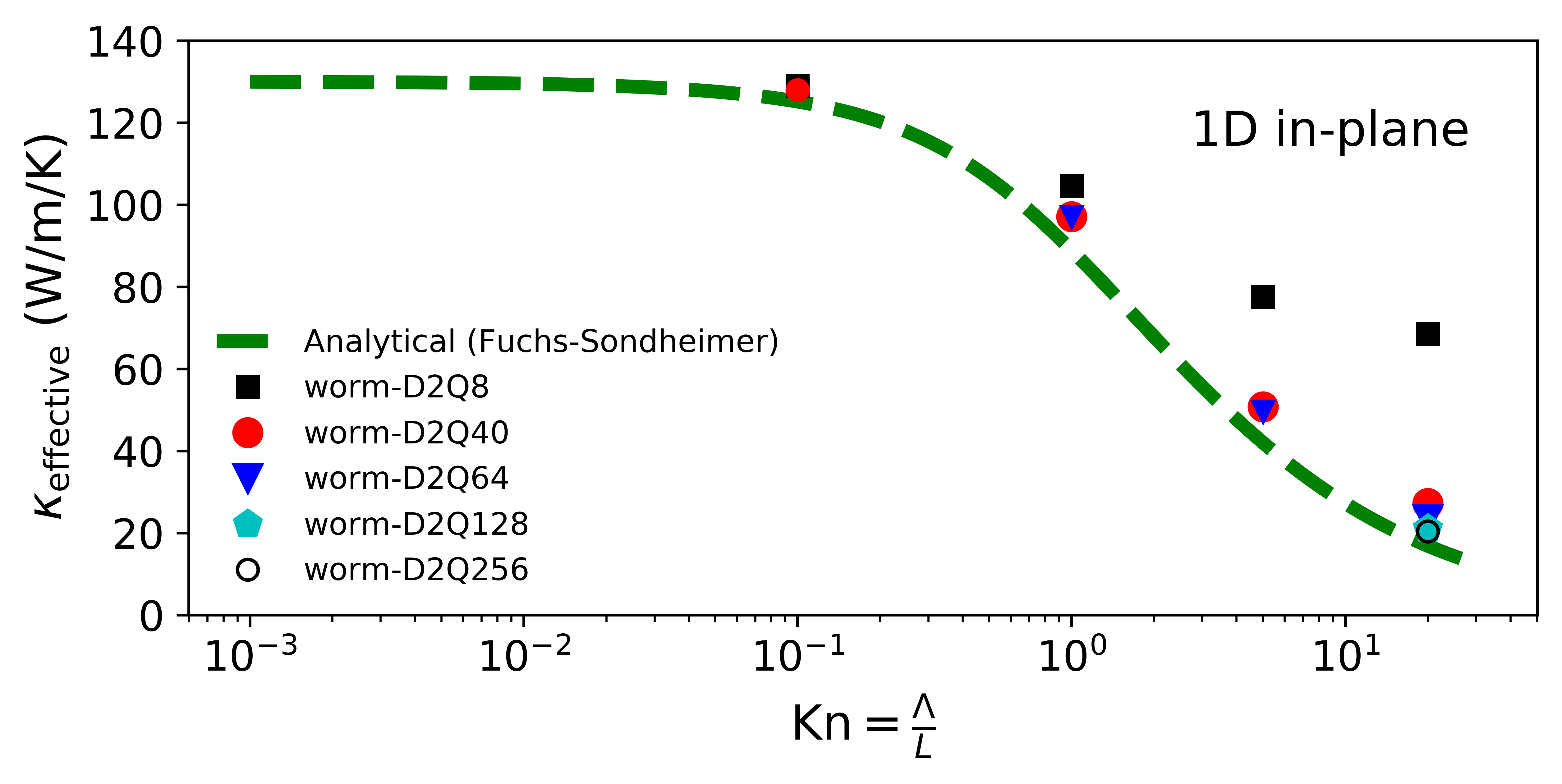}
    \caption{ Knudsen number dependence of the 1D in-plane thermal conductivity at     300.25 K. The data correspond to: $N_x \times N_y  =318\times318$, and $t/\Delta t=160000$ (Kn=0.1);   $N_x \times N_y  =5118\times38$, and $t/\Delta t=256000$ (Kn=1, 5). For Kn=20 the data  corresponds to $t/\Delta t=64000$. The numbers reported inside the graph correspond to the error in percentage of the worm-LBM numerical solution, taking as  reference the analytical value. }
    \label{fig:kappa-inplane}
\end{figure}

Taking into account the convergence tests in terms of number of directions, grid size, and length-width ratios, the Knudsen number dependence of the in-plane effective thermal conductivity was compared to the analytical solution (Fig. \ref{fig:kappa-inplane}). As it can be seen, LBM schemes with high number of propagation directions are necessary to correctly describe the in-plane thermal conductivity in the ballistic regime. The standard scheme with only eight directions results in a considerable overestimation of the analytical solution (e.g. 307 \% for Kn=20).  As explained before, the problem has its origin in a too large proportion of energy propagating along the axial direction, which never interacts with the adiabatic diffuse  wall. The worm-LBM naturally remedies the problem by allowing more propagation directions, without having to resort in fitting procedures to the analytical solution as previously proposed by other authors \cite{GUO20161}.  As it can be seen in Fig. \ref{fig:kappa-inplane}, the numerical solution converges with the number of propagation directions toward the analytical value. For the intermediate and low Knudsen numbers (Kn = 0.1, 1, 5), the effective thermal conductivity does not change further for Q>40. Conversely, for Kn=20, the values calculated with Q=128 and Q=256 still differ in around $\sim$ 4\%. In all cases, finer grids would be necessary to obtain a better agreement with the analytical solution. Here we have omitted such grid convergence, as the main objective of this test case was to show the impact of the propagation directions.

\subsection{The stationary one hot - three cold boundaries}
\label{sec:stationary-one-hot-three-cold}
\begin{figure}[h]
    \centering
    \includegraphics[width=0.48\textwidth]{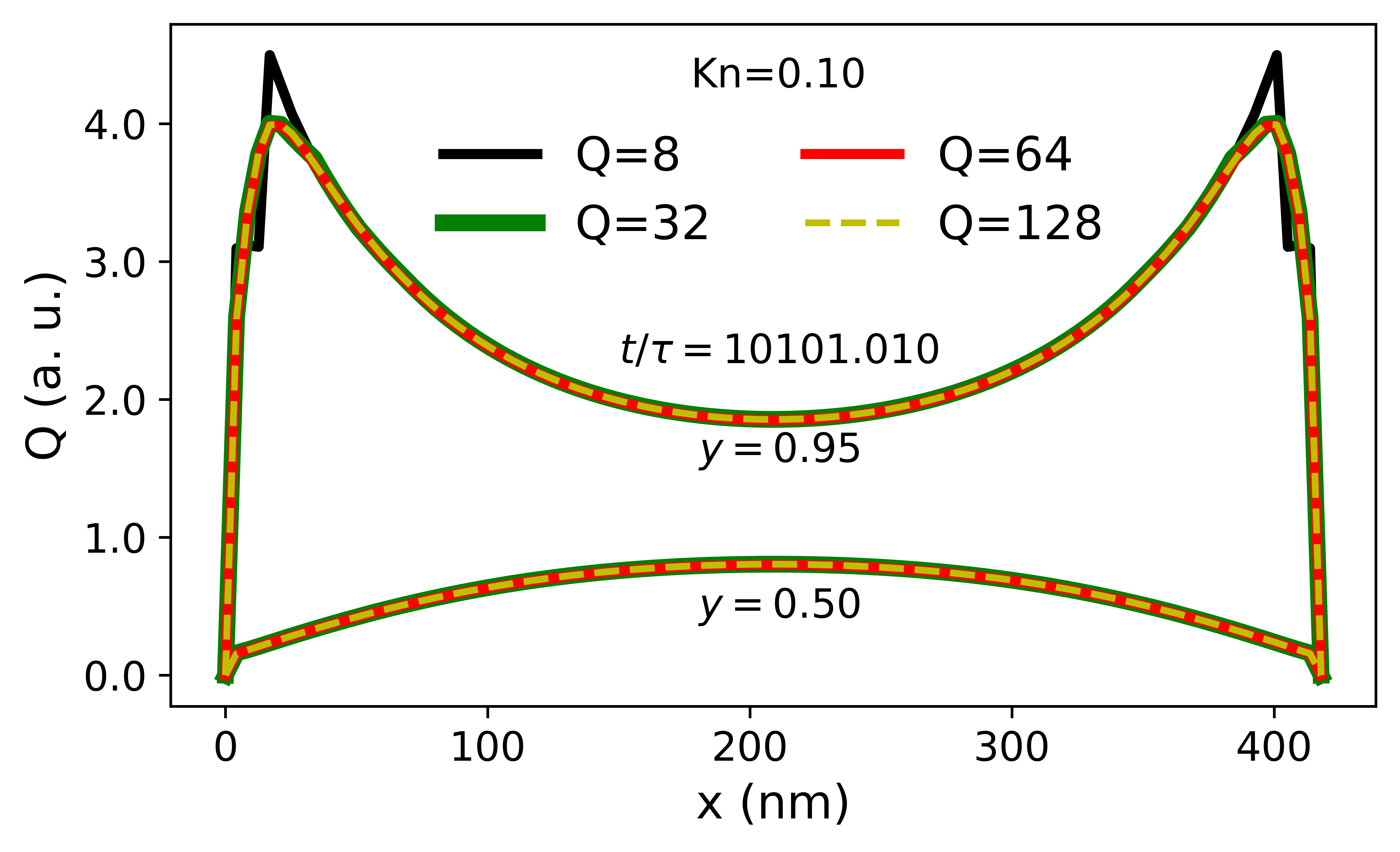}
    \caption{$Q$-dependence of the  heat flux profile for $\text{Kn}=0.1$ . The  profiles were calculated along the $X$ directions for $Y$ = 0.5$\,l_y$, 0.75$\,l_y$, and 0.95$\,l_y$, where $l_y$ is the length of the domain along the $Y$ direction ($l_x=l_y=l$). A grid of $N_x \times N_y= 98 \times 98$ was used. }
    \label{fig:2D-onehot-threecold-Q}
\end{figure}

\begin{figure}[h]
    \centering
    \includegraphics[width=0.5\textwidth]{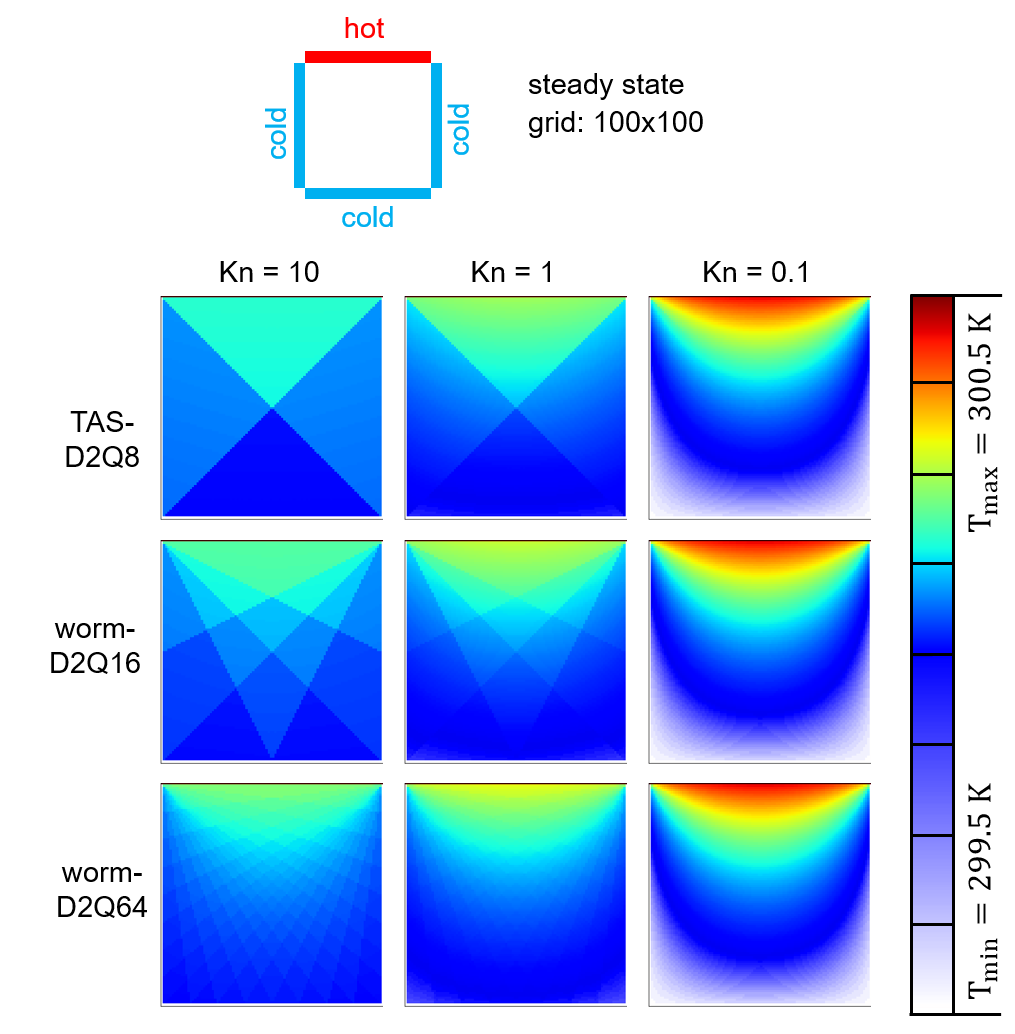}
    \caption{Steady state temperature distribution for different Knudsen numbers Kn (0.1, 1, 10), and Q directions (8, 16, 64), as calculated by different LBM schemes. }
    \label{fig:2D-onehot-threecold-Kn}
\end{figure}

The one hot - three cold boundary condition problem  consists of four isotropic blackbody walls  at  temperatures  $T_h=300.5$ K (upper wall) and  $T_c=299.5$ K (remaining walls). A Knudsen number of $\text{Kn}=0.005$ is imposed for the propagation from the upper to the lower wall. 
\begin{figure}[h]
    \centering
    \includegraphics[width=0.48\textwidth]{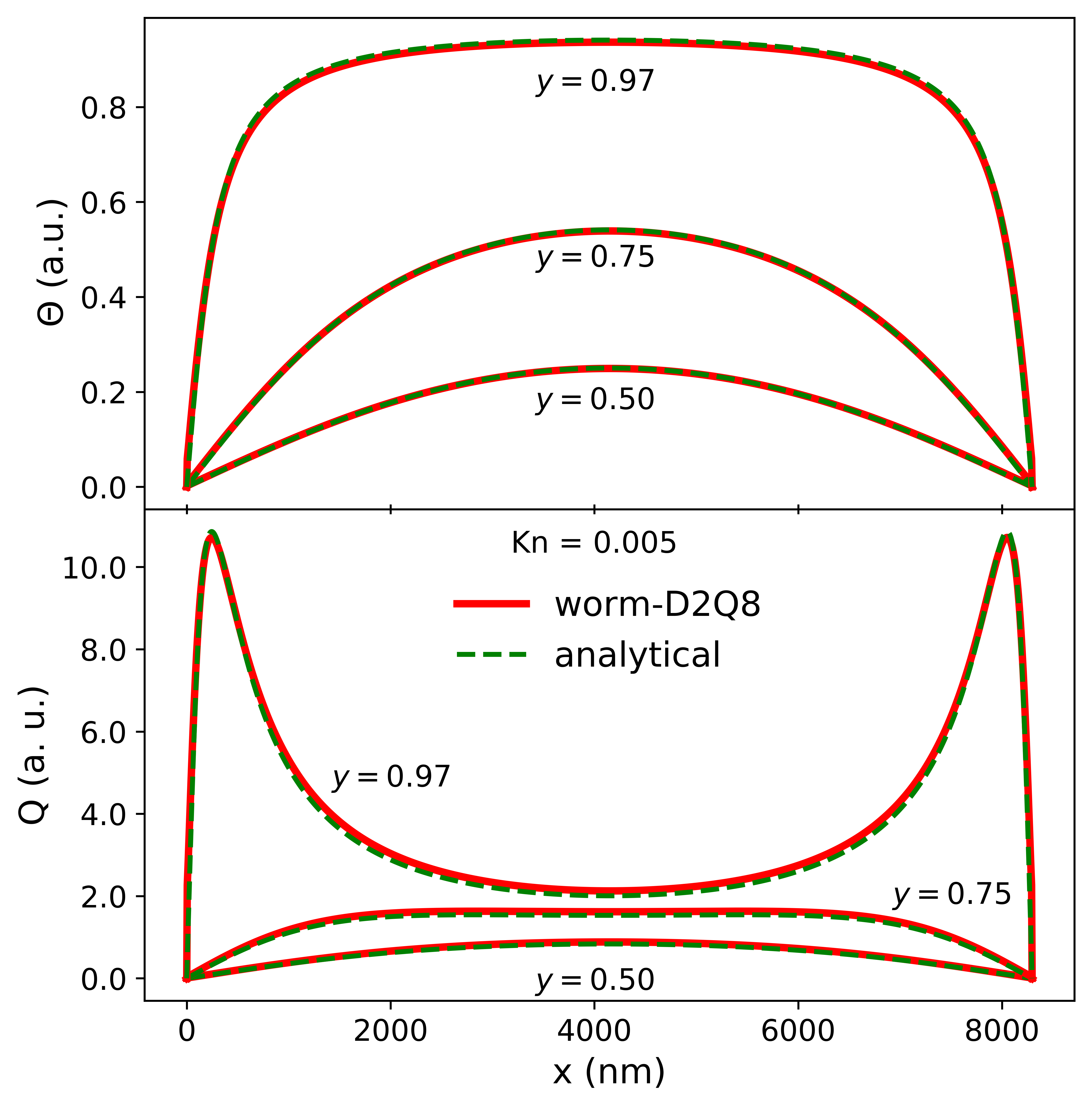}
    \caption{ worm-D2Q8 steady state temperature and heat flux profiles compared to the analytical solutions  (Eqs. \ref{eq:2DT} and \ref{eq:2DTy}). The  profiles were calculated along the $X$ directions for $Y$ = 0.5$\,l_y$, 0.75$\,l_y$, and 0.97$\,l_y$, where $l_y$ is the length of the domain along the $Y$ direction ($l_x=l_y=l$). A grid of $N_x \times N_y= 1598 \times 1598$ was used.}
    \label{fig:2D-onehot-threecold-Fourier}
\end{figure}
In the diffusive limit, the temperature and heat flux profiles in the steady state along the $X$ direction at different $Y$ heights were calculated with the worm-D2Q8 scheme (Fig. \ref{fig:2D-onehot-threecold-Fourier}), and compared to the analytical solutions: 
\begin{equation}
    \Theta(X,Y) =  \sum_{l=1}^\infty \frac{2[1-\cos{(l\,\pi)}]}{l\pi \sinh{(l\,\pi)}}\sin{(l\,\pi\, X)}\sinh{(l\pi Y)},
    \label{eq:2DT}
\end{equation}
\begin{equation}
    Q_x(X,Y) =  -\sum_{l=1}^\infty \frac{2[1-\cos{(l\,\pi)}]}{ \sinh{(l\,\pi)}}\cos{(l\,\pi X)}\sinh{(l\,\pi Y)},
    \label{eq:2DQx}
\end{equation}
\begin{equation}
    Q_y(X,Y) =  -\sum_{l=1}^\infty \frac{2[1-\cos{(l\,\pi)}]}{ \sinh{(l\,\pi)}}\sin{(l\,\pi\, X)}\cosh{(l\,\pi\, Y)},
    \label{eq:2DTy}
\end{equation}
with $X = \frac{x}{L}$ and $Y = \frac{y}{L}$.  
 Overall,  the heat flux and temperature profiles within the domain match the analytical Fourier solution very well. However, the agreement gets reduced as the distance to the hot boundary  becomes  smaller (e.g. $0.97\,l_y$).  Close to this boundary, the transport problem becomes more ballistic, and the Fourier solution is not valid anymore. Consistently, we found that schemes with a low number of propagation  directions  (e.g. D2Q8) are less accurate than higher-direction schemes in the vicinity of the hot boundary, and particularly close to the corners where ballistic transport dominates. Figure  \ref{fig:2D-onehot-threecold-Q} shows a comparison of the transient heat flux as obtained with different worm-D2QM (M=8, 32, 64, and 128) schemes. In regions where diffusive transport dominates, the numerical solution is independent of the number of propagation directions (i.e. Q), while it becomes dependent in the ballistic areas.
The impact of the number of propagation directions is, of course, higher as the entire domain becomes dominated by ballistic transport. In the limit of high Knudsen numbers ($\text{Kn} = 1$ and $10$),  schemes with a too low number of propagation directions (e.g. $Q=8$ and $16$) suffer much more from the ray effect problem than domains dominated by small Knudsen numbers ($\text{Kn}=0.1$), see Fig. \ref{fig:2D-onehot-threecold-Kn}.

\begin{figure*}
    \centering
    \includegraphics[width=1.0\textwidth]{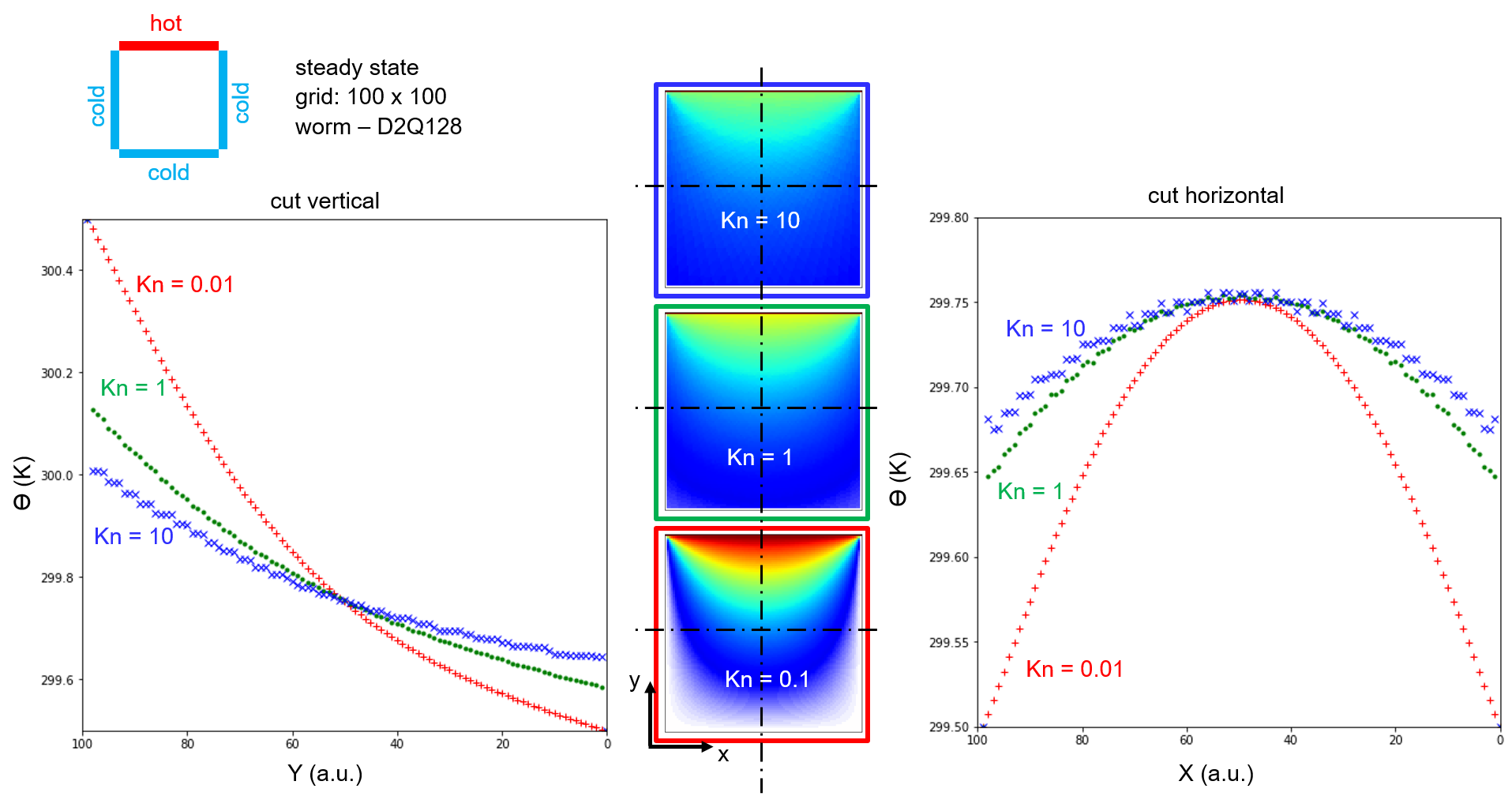}
    \caption{worm-D2Q128 steady state temperature profiles  for $\text{Kn} = 0.01, 1$ and $10$. Left panel: temperature profile along the cross section at $X=1/2$.  Right panel: temperature profile along the cross section at $Y=1/2$.}
    \label{fig:2D-onehot-threecold-ballistic}
\end{figure*}

A clear difference between the transport character of this problem in the ballistic and diffusive regime can be seen from comparing the temperature profiles for different Knudsen numbers (Fig. \ref{fig:2D-onehot-threecold-ballistic}).  As  the Knudsen number increases, a temperature slip at the limits of the domain is observed. The larger the jump, the more the ballistic transport dominates. 

\section{Conclusions}

In this work, the simple and powerful worm-LBM for solving Boltzmann-type transport equations is proposed. It is an improved lattice Boltzmann method that does not suffer from the ray effect problem, which so far has rendered this method inaccurate to describe systems with high Knudsen numbers correctly (caused by a limited number of propagation directions). The worm-LBM is a simple algorithm to implement multiple propagation directions within a square scheme of the type of  D2Q[$M\times$8] ($M\geq1$), at the same computational cost per direction as in a standard D2Q8 scheme. The worm-LBM makes also use of an adaptive time scheme (TAS) that allows to implement isotropic as well as angular dependent propagation speeds on a regular square grid. The maximum deviation from a desired velocity and propagation direction during a worm-LBM simulation is less or equal to one grid spacing at any time. This makes the method particularly powerful for long time simulations.

The numerical accuracy  of the worm-LBM algorithm was verified in the framework of thermal transport in the gray approximation.  Its suitability for describing ballistic and diffusive phonon transport problems in transient and steady states was demonstrated. For the full range of Knudsen numbers and test cases, including initial and boundary conditions, the worm-LBM has shown a first-order convergence rate. The modest convergence rate is balanced by the fact that the worm-LBM already provides accurate results with coarse grids. The largest errors are observed for the coarsest possible grid (one computational cell per mean free path) at the lowest simulated Knudsen numbers (Kn=0.005-0.01). The relative error in temperature for the cases reported at these Knudsen numbers was below 0.01 \%. Conversely, for high Knudsen numbers (e.g. Kn=20), which imply a higher number of computational cells per mean free path, even the coarsest grid used resulted in relative errors for the temperature below 0.0001 \%.

It was shown that challenging problems as the in-plane thermal transport in the ballistic regime are correctly accounted for by the new worm-LBM. The in-plane thermal transport problem, from intermediate to high Knudsen numbers, serves as a representative example of general geometries where a sufficiently high resolution in angular space is required (i.e.  high number of propagation directions is needed). 

The reliability of the new worm-LBM to describe phonon thermal transport, so far, has been shown in the limit of the gray approximation. Future work will involve a benchmark of the method for a full phonon dispersion, using as an input phonon properties obtained from ab initio calculations.  In this regard, the algorithm has the additional advantage that different velocities can be simulated on the same grid. 
 
Overall, due to its efficiency, simplicity, and advantages  (i.e. reduced ray effect, no numerical smearing or angular false scattering), the worm-LBM has the potential of becoming the forefront methodology to tackle transport processes in a wide variety of fields.
\section{Acknowledgements}
The authors gratefully acknowledge the financial support under the scope of the COMET program within the K2 Center “Integrated Computational Material, Process and Product Engineering (IC-MPPE)” (Project No 859480). This program is supported by the Austrian Federal Ministries for Climate Action, Environment, Energy, Mobility, Innovation and Technology (BMK) and for Digital and Economic Affairs (BMDW), represented by the Austrian research funding association (FFG), and the federal states of Styria, Upper Austria and Tyrol.

\section{Appendix A. Stability and dispersion relation analysis}
The  worm-LBM is described by linear equations (see Eq. \ref{eq:algorithm}). Therefore, its  stability  can be checked by using the  von Neumann stability analysis. For simplicity we assume homogeneous $W$ and initial conditions, which allow to write $e_i^\text{eq}=e_i=e(x_i)$.  An algorithm is stable if the growth factor $|g|\leq 1$ for:
\begin{equation}
   \tilde{u}^{t+\Delta t}(k)=g(k) \;\tilde{u}^t(k) \quad k\in \left[-\frac{\pi}{\Delta x},\frac{\pi}{\Delta x}\right], 
\end{equation}
where $\tilde{u}(k)$ is the Fourier transform of $u(x_i)=e(x_i)$. For a step where all directions are propagating one gets: 
\begin{align}
    \tilde{u}^{t+\Delta t} &= (1-W) e^{i k \Delta x} \tilde{u}^{t}+ W e^{i k \Delta x} \tilde{u}^{t}\nonumber \\
    \tilde{u}^{t+\Delta t}&= e^{i k \Delta x} \tilde{u}^{t} \rightarrow |g| = 1,
    \label{eq:stabil-prop}
\end{align}
thus, this step is stable for all $W$. For a step where the propagation of the DDEDs is paused one obtains:
\begin{align}
    \tilde{u}^{t+\Delta t} &= (1-W) \tilde{u}^{t} + W e^{i k \Delta x} \tilde{u}^{t}\nonumber \\
    \tilde{u}^{t+\Delta t} &= \left[1 + W\left(e^{i k \Delta x}-1\right)\right] \tilde{u}^{t}.
    \label{eq:stabil-pause}
\end{align}
In the limit of $W=0$ and $W=1$ the growth factor is exactly $1$. In general, in the complex plane this result constitutes a circle with radius $W$, with its midpoint shifted along the real axis by $1-W$. Therefore, the value with the highest and lowest magnitude are located on the real axis.
The real part of $g$ is:
\begin{equation}
\text{Re}(g) = 1 + W \left[\,\cos (k \Delta x) - 1\,\right],
\end{equation}
showing a minimum of $1-2\,W$ and a maximum of $1$. As a result the magnitude of the growth factor is smaller or equal to one for $W\leq1$, which provides the stability limit for the scheme.

The dispersion relation is obtained by performing the Fourier transform also in time, where for the case of propagating DDESs one gets:
\begin{align}
    e^{i \omega \Delta t} \tilde{u} &= e^{i k \Delta x} \tilde{u} \nonumber \\
    \omega &= \frac{\Delta x}{\Delta t } k = c\,k,
    \label{eq:disp-prop}
\end{align}
where $c$ is the magnitude of the lattice velocity $\vec{c}_i$. For steps where the DDED propagation is paused one obtains:
\begin{align}
    e^{i \omega \Delta t} \tilde{u} &= 1 + W(e^{i k \Delta x} -1) \tilde{u} \nonumber \\
    \omega &=  -\frac{i}{\Delta t}\ln\left[1 + W(e^{i k \Delta x} -1)\right].
    \label{eq:disp-pause}
\end{align}
Thus for $W=0$ one obtains $\omega=0$, corresponding a non-changing solution, whereas for $W=1$ one obtains the linear dispersion $\omega=c\,k$. 
\section{References}
\bibliographystyle{elsarticle-num}
\bibliography{LBE.bib}


\end{document}